\pdfoutput=1
\documentclass[12pt,preprint]{aastex}

\usepackage{graphicx}
\usepackage{chngpage}

\newcommand{\et}{et al.}
\newcommand{\kms}{km s$^{-1}$}
\newcommand{\ha}{H$\alpha$}
\newcommand{\solar}{\ifmmode_{\sun}\;\else$_{\sun}$\fi}

\newcommand{\HII}{H$\,${\sc ii}}
\newcommand{\HI}{H$\,${\sc i}}

\begin{document}

\title{LITTLE THINGS}

\author{Deidre A. Hunter\altaffilmark{1}, Dana Ficut-Vicas\altaffilmark{2},
Trisha Ashley\altaffilmark{3}, 
Elias Brinks\altaffilmark{2}, Phil Cigan\altaffilmark{4}, Bruce G. Elmegreen\altaffilmark{5}, 
Volker Heesen\altaffilmark{2,6},  Kimberly A. Herrmann\altaffilmark{1,7},
Megan Johnson\altaffilmark{1,8,9}, Se-Heon Oh\altaffilmark{10,11}, Michael P. Rupen\altaffilmark{12}, 
Andreas Schruba\altaffilmark{13,14}, Caroline E. Simpson\altaffilmark{3},
Fabian Walter\altaffilmark{13}, David J. Westpfahl\altaffilmark{4}, Lisa M. Young\altaffilmark{4}, and
Hong-Xin Zhang\altaffilmark{1,15,16}}

\altaffiltext{1}{Lowell Observatory, 1400 West Mars Hill Road, Flagstaff, Arizona 86001 USA}

\altaffiltext{2}{Centre for Astrophysics Research, University of Hertfordshire, College Lane, Hatfield, AL10 9AB
United Kingdom}

\altaffiltext{3}{Department of Physics, Florida International University, CP 204, 11200 SW 8th St,
Miami, Florida 33199 USA}

\altaffiltext{4}{Physics Department, New Mexico Institute of Mining and Technology, Socorro, New Mexico 87801
USA}

\altaffiltext{5}{IBM T. J. Watson Research Center, PO Box 218, Yorktown Heights, New York 10598 USA}

\altaffiltext{6}{Current address: School of Physics and Astronomy, University of Southampton, Southampton SO17 1BJ UK}

\altaffiltext{7}{Current address: Department of Physics, Pennsylvania State University Mont Alto, Science
Technology Building, Mont Alto, PA 17237 USA}

\altaffiltext{8}{Department of Physics and Astronomy, Georgia State University, 29 Peachtree Center Avenue,
Science Annex, Suite 400, Atlanta, Georgia 30303-4106 USA}

\altaffiltext{9}{Current address: National Radio Astronomy Observatory, PO Box 2, Green Bank, WV 24944-0002
USA}

\altaffiltext{10}{The International Centre for Radio Astronomy Research (ICRAR), The University of Western Australia,
35 Stirling Highway, Crawley, Western Australia 6009, Australia}

\altaffiltext{11}{ARC Centre of Excellence for All-sky Astrophysics (CAASTRO)}

\altaffiltext{12}{National Radio Astronomy Observatory, 1003 Lopezville Road, Socorro, NM 87801 USA}

\altaffiltext{13}{Max-Planck-Institut f\"ur Astronomie, K\"onigstuhl 17, 69117 Heidelberg Germany}

\altaffiltext{14}{Current address: Center for Astronomy and Astrophysics, California Institute of Technology,
1200 E California Blvd, MC 249-17, Pasadena, CA 91125, USA}

\altaffiltext{15}{Purple Mountain Observatory, Chinese Academy of Sciences, 2 West Beijing Road, 
Nanjing 210008 P. R. China}

\altaffiltext{16}{Current address: Peking University, Astronomy Department, No. 5 Yiheyuan Road, Haidian District, 
Beijing, P. R. China 100871}

\begin{abstract}
We present LITTLE THINGS (Local Irregulars That Trace Luminosity Extremes, 
The \HI\ Nearby Galaxy Survey) that is aimed at determining what drives star formation in dwarf galaxies. 
This is a multi-wavelength survey of 37 Dwarf Irregular and 4 Blue Compact Dwarf galaxies that
is centered around \HI-line data obtained with the National Radio Astronomy Observatory
(NRAO) Very Large Array (VLA).
The \HI-line data are characterized by high sensitivity 
($\leq1.1$ mJy beam$^{-1}$ per channel), high spectral resolution ($\leq$2.6 \kms),
and high angular resolution ($\sim$6\arcsec).
The LITTLE THINGS sample contains dwarf galaxies that are relatively nearby ($\leq$10.3 Mpc; 6\arcsec\ is $\leq$300 pc), 
that were known to contain atomic hydrogen, the fuel for star formation, and that cover a large range in dwarf galactic properties.
We describe our VLA data acquisition, calibration, and mapping procedures, as well as \HI\ map characteristics,
and show channel maps, moment maps, velocity-flux profiles, and surface gas density profiles.
In addition to the \HI\ data we have {\it GALEX} UV and ground-based $UBV$ and \ha\ images for most
of the galaxies, and $JHK$ images for  some. {\it Spitzer} mid-IR images are available for many of the
galaxies as well. 
These data sets are available on-line.
\end{abstract}


\section{Introduction} \label{sec-intro}

Dwarf Irregular (dIrr) galaxies are the most common type of galaxy in the local universe (see, for example, Dale \et\ 2009). 
They are also the closest analogs
to the low mass dark matter halos that formed after the Big Bang.
In the $\Lambda$CDM model, it is in these entities that the
first stars formed. Yet, we do not understand what drives
star formation on galactic scales even in nearby dwarfs, the simplest, most pristine local environments
(see, for example, Hunter 2008, Bigiel \et\ 2010).
To remedy this situation we have assembled multi-wavelength data on a large sample of relatively
normal, nearby gas-rich dwarf galaxies, tracing their stellar populations, gas content, dynamics,
and star formation indicators.
This project is called LITTLE THINGS (Local Irregulars That Trace Luminosity Extremes, 
The \HI\ Nearby Galaxy Survey), and it builds on the THINGS project, whose emphasis was on nearby
spirals (Walter \et\  2008). 

LITTLE THINGS\footnote[17]{http://www.lowell.edu/users/dah/littlethings/index.html}
was granted $\sim$376 hours with the Very Large Array 
(VLA\footnote[18]{
The VLA is a facility of the National Radio Astronomy Observatory (NRAO).
The NRAO is a facility of the National Science Foundation operated under cooperative agreement by Associated Universities, Inc.
These data were taken during the upgrade of the VLA to the Expanded VLA or EVLA. 
In this paper we refer to the instrument as the VLA, the retrofitted antennae as EVLA antennae, and non-retrofitted antennae
as VLA antennae. This emphasizes the hybrid nature of the instrument and distinguishes it from the far more powerful Jansky VLA or
JVLA it has become since 2012.
})
in B, C, and D configurations to obtain \HI-line data of 21 dIrr and Blue Compact Dwarf (BCD) galaxies.
We added another 20 galaxies from the VLA archives for a total sample of 37 
dIrr and 4 BCD galaxies covering a large range of dwarf galactic parameter space.
Our VLA data emphasize deep, high spatial (6\arcsec) and spectral ($\le$2.6 km s$^{-1}$) resolution
in order to reveal clouds, shells, and turbulence in the interstellar medium (ISM) 
that could be important for creating regions of higher density star-forming gas.
We are combining the \HI-line data with 
ultraviolet (UV), optical, and infrared (IR) data to address the following questions:

{\it What regulates star formation in small, gas-rich galaxies?}
Studies of the star formation laws in galaxies show that star formation is very inefficient in
dwarf galaxies and in the outer disks of spirals (Bigiel \et\ 2010; Ficut-Vicas \et, in preparation).
Furthermore, the star formation efficiency decreases as the gas density decreases.
How then do atomic hydrogen clouds form in the low gas density environments of dwarfs
that ultimately lead to star-forming molecular clouds?

{\it What is the relative importance of sequential triggering by previous generations of stars
for star formation in dwarf galaxies?}
One generation of stars can trigger the formation of the next by rearranging the gas through 
winds and supernova explosions (see, for example, \"Opik 1953, Gerola \et\ 1980, Comins 1983, Dopita \et\ 1985,
Mori \et 1997, van Dyk \et\ 1998), 
but how important is this process? 
Observations of dwarfs show a better correlation between the star formation rate and the $V$-band surface 
brightness, which emphasizes Gyr old stars, than with any other measure, 
including the average gas density (Hunter \et\ 1998). 
Does this relationship imply 
a significant causal relationship in which existing stars are important for triggering new star formation?

{\it What is the relative importance of random turbulent compression cloud formation in dwarf galaxies?}
Turbulence can account for a wide range of phenomena that are indirectly related to star formation:
the power law luminosity functions of \HII\ regions, the power law mass functions of clouds and clusters 
(Mac Low \& Klessen 2004), the shape of the power spectra of \HI\
(Stanimirovi\'c \et\ 1999, Elmegreen \et\ 2001, Dickey \et\ 2001, Zhang \et\ 2012a) and H$\alpha$
(Willett \et\ 2005), the log-normal distribution of the H$\alpha$ intensity in individual pixels in H$\alpha$ images of dwarf galaxies,
(Hunter \& Elmegreen 2004), the correlation between region size and the star formation time scale
(Efremov \& Elmegreen 1998), and hierarchical structures in the ISM (Bastian \et\ 2007).
But how important is turbulence in the formation of star-forming clouds, particularly in the realm of sub-critical gas density?

{\it What happens to the star formation process in the outer parts of disks?}
We find that many dwarfs have exponential stellar surface brightness profiles with a break and a change in slope. 
These breaks are similar to what are seen in the outer parts of spirals
(e.g., de Grijs \et\ 2001; Kregel \et\ 2002; Pohlen \et\ 2002; Kregel \& van der Kruit 2004; 
Pohlen \& Trujillo 2006; Erwin \et\ 2008; Guti\'{e}rrez \et\ 2011)
and disks at high redshift (P\'erez 2004).
The break could imply a change in the star formation process at the break radius,
but what is really happening there?  
  
Here we describe the LITTLE THINGS project and present the \HI\ data. Science papers that are part of
this project include: 
1) ``Velocity Dispersions and Star Formation Rates in the LITTLE THINGS Dwarf Galaxies (Cigan \et, in preparation),
2) ``Star Formation Laws in LITTLE THINGS Dwarfs: The case of DDO 133 and DDO 168'' (Ficut-Vicas \et, in preparation),
3) ``Deep Radio Continuum Imaging of the Dwarf Irregular Galaxy IC 10: Tracing Star Formation and Magnetic Fields''
(Heesen \et\ 2011),
4) ``Deep Radio Continuum Observations of the Nearby Dwarf Irregular Galaxy IC 10. I. Nature of the Radio Continuum Emission'' (Heesen \et, in preparation),
5) ``Surface Brightness Profiles of Dwarf Galaxies. I. Profiles and Statistics'' (Herrmann \et\ 2012),
6) ``Surface Brightness Profiles of Dwarf Galaxies. II. Color Trends and Mass Profiles'' (Herrmann \et, in preparation),
7) ``The Stellar and Gas Kinematics of the LITTLE THINGS Dwarf Irregular Galaxy NGC 1569'' (Johnson \et\ 2012),
8) ``The Stellar Kinematics of the LITTLE THINGS Dwarf Irregular Galaxies DDO 46 and DDO 168'' (Johnson \et, in preparation),
9) ``Deep 6 cm Radio Continuum Observations of the Nearby Dwarf Irregular Galaxy IC 1613'' (Kitchener \et, in preparation),
10) ``High-resolution Mass Models of Dwarf Galaxies from LITTLE THINGS'' (Oh \et, in preparation),
11) ``Star Formation in IC 1613" (Simpson \et, in preparation),
12) ``Outside-in Shrinking of the Star-forming Disk of Dwarf Irregular Galaxies" (Zhang \et\ 2012b),
and 13) ``\HI\ Power Spectrum Analysis of Dwarf Irregular Galaxies'' (Zhang \et\ 2012a).
Other science that is in progress includes, but is not limited to, studies of the following:
1) Star Formation Processes in Blue Compact Dwarfs -- Ashley,
2) Molecular Cloud Structure and Dust at Low Metallicity -- Cigan,
3) Star Formation Laws of the full sample of galaxies -- Ficut-Vicas,
4) Surface Brightness Profiles and their relationship to \HI\ gas and two dimensional images -- Herrmann,
5) The Role of Turbulence in Star Formation in Dwarfs -- Hunter,
6) A Deep 6 cm Radio Continuum Survey of LITTLE THINGS -- Kitchener,
and 7) Porosity of the Interstellar Medium in Dwarfs -- Simpson.
   
\section{The Survey} \label{sec-survey}

\subsection{Sample selection}

The LITTLE THINGS sample of 37 dIrr galaxies and 4 BCDs
was drawn from a larger multi-wavelength survey of 94
relatively nearby dIrr galaxies and 24 BCD systems, as well as 18 Sm galaxies (Hunter \& Elmegreen 2004, 2006).
A large sample is necessary to disentangle complex processes in
surprisingly complicated galaxies. This larger sample covers nearly the full range of galactic
parameters seen in dwarf galaxies including 
luminosity (9 magnitudes in absolute magnitude M$_V$), 
current star formation rate normalized to the area within one disk scale length (a range of a factor of $10^4$),
relative gas content (a factor of 110 in $M_{HI}/L_B$), and
central surface brightness (7 magnitudes in $\mu_0^V$, the central $V$-band surface brightness).
The sample does not include obviously interacting
galaxies because we are investigating internal processes, and
it only includes dwarfs with substantial amounts of atomic hydrogen so that they {\it could} be forming stars.

\begin{figure}
\epsscale{1.0}
\includegraphics[angle=0,width=0.8\textwidth]{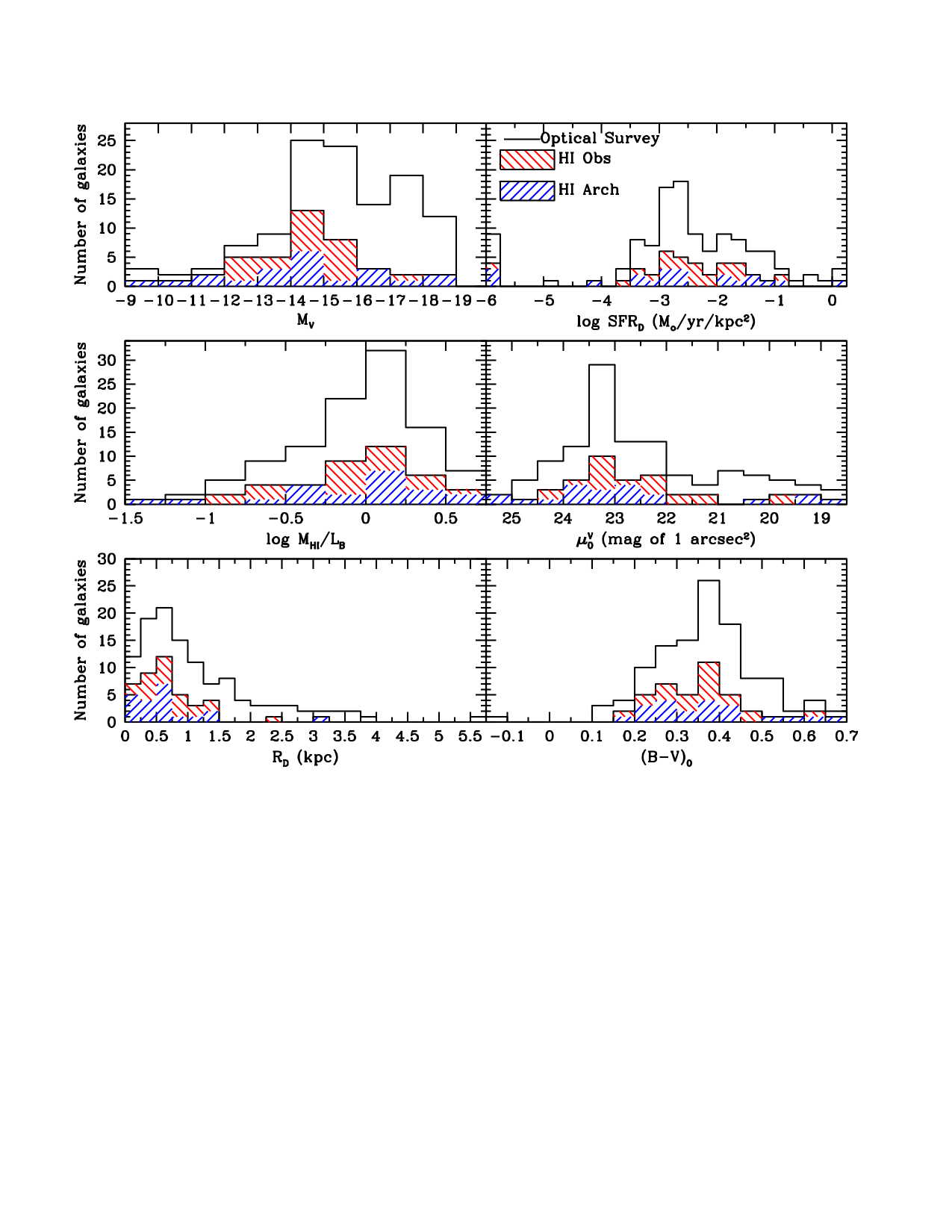}
\caption{Properties of the LITTLE THINGS sample
(new VLA observations [{\it hashed $+45^\circ$}] and archive data [{\it hashed $-45^\circ$}])
compared to the entire optical survey of Hunter \& Elmegreen (2006; {\it no hash}).
The LITTLE THINGS sample covers the range of parameters of the full survey.
SFR$_D$ is the star formation rate, based on H$\alpha$ emission, normalized to $\pi R_D^2$, where
$R_D$ is the disk scale length determined from a $V$-band image.
A star formation rate per unit area of 0 is plotted as a log of $-6$. 
The parameter $\mu_0^V$ is the central surface brightness determined from a fit to the 
$V$-band surface brightness profile.
The distance dependent parameters M$_V$ and $R_D$ use the distances given in Table \ref{tab-sample}
for the LITTLE THINGS sample and the distances given by Hunter \& Elmegreen (2006) for the rest
of the galaxies in the larger optical survey sample.
\label{fig-props}}
\end{figure}

For LITTLE THINGS, we excluded galaxies with distances $>$10 Mpc in order to achieve
a reasonably small spatial resolution in the VLA \HI\ maps
($<$300 pc at 10 Mpc in B array), although the distance to
our most distant galaxy was revised later to 10.3 Mpc.
The average distance to the galaxies in our sample is 3.7 Mpc.
In addition we excluded galaxies with
$W_{20}>$160 \kms\ in order to fit the galaxy emission comfortably in the bandwidth available for our desired velocity resolution,
leaving room for determining continuum emission.
The parameter $W_{20}$ is the full-width at 20\% of the peak of an integrated \HI\ flux-velocity profile (Bottinelli \et\ 1990).
One galaxy originally in our sample, NGC 1156, was eliminated when we discovered that the archival
data contained no line-free channels.
We also excluded galaxies with high inclinations ($>$70$^{\circ}$) so that we can better resolve structures in the disks.
We then insisted that the LITTLE THINGS sample cover the range of
integrated properties seen in the full optical survey, especially
$M_V$ and star formation rate, including
galaxies without any measurable \ha\ emission.
A comparison of the properties of the LITTLE THINGS sample with the larger parent sample
is shown in Figure \ref{fig-props}.
Our sample includes 19 dIrr galaxies and one BCD galaxy in the VLA
archives having B/C/D-array observations at 2.6 \kms\ channel separation or better.
Our new VLA observations are of 18 dIrr galaxies and 3 BCDs in the B/C/D arrays.
In total, 19 galaxies have \HI-line data at 1.3 \kms\ channel separation.

One of our 41 galaxies, NGC 6822, which we observed at 1.3 \kms\ channel separation,
is highly extended compared to the primary beam.
Those data require special treatment and calibration and mapping are in progress.
Below, we list the data we obtained on NGC 6822, 
but we do not include it in the discussions of the \HI\ maps and cubes.

Table \ref{tab-sample} lists the sample galaxies and several basic properties, including other names
by which the galaxies are known, distances, $M_V$,  a measure of size, foreground reddening,
star formation rates, and oxygen abundance.
Here we elaborate on a few of the quantities:
1) Distance. In choosing a distance for each galaxy, we considered the references for 
distance determinations given by the NASA/IPAC Extragalactic Database (NED).
In order of preference, we chose distances determined from Cepheids or RR Lyrae stars and from the tip of the
red giant branch. The references for the distance used for each galaxy are given in Table \ref{tab-sample}. 
If none of those determinations was available, we used the Hubble flow distance
determined from the galaxy's radial velocity corrected for infall to the Virgo Cluster 
(Mould \et\ 2000) and a Hubble constant of 73 km s$^{-1}$ Mpc$^{-1}$. 
2) Size measurement. For a measure of the size of the galaxy, we used the Holmberg radius, $R_H$, 
which was originally measured by Holmberg (1958) at a photographic surface
brightness of 26.5 mag in 1 arcsec$^2$. Here we have used the $R_H$ measured
by Hunter \& Elmegreen (2006), who converted from
photographic magnitudes to $B$-band magnitudes using the recipe of
de Vaucouleurs \et\ (1976). For our galaxies this corresponds to a $B$-band surface
brightness of between 26.60 and 26.72 mag in 1 arcsec$^2$ with
an average of 26.64$\pm$0.002 mag in 1 arcsec$^2$. 
3) Star formation rate (SFR). Star formation rates are taken from Hunter \& Elmegreen (2004). These were
determined from integrated \ha\ luminosities and are normalized to the area encompassed by
one disk scale length. We use this characteristic area to normalize the SFRs
in order to compare galaxies of different size, but this is not the area that encompasses \ha\ emission.
Some galaxies have \ha\ emission that extends less than and others more than one disk scale length.
4) Oxygen abundances. As a measure of the metallicity of the galaxy we list oxygen abundances, 
$12+\log {\rm O/H}$, taken from the literature and determined from \HII\ regions. 
For galaxies without \HII\ region abundance measurements, we used $M_B$ and the luminosity-metallicity
relationship of Richer \& McCall (1995). The abundance determined this way, listed in parentheses, is very uncertain, 
but it gives an indication of the expected relative metallicity of the galaxy.
The references for the abundances are given in the Table. 
For comparison the solar oxygen abundance value is 8.69 (Asplund \et\ 2009).      
      
\subsection{VLA observational philosophy}

Our new VLA observations are characterized by deep observations with high angular and high spectral resolution.
To achieve the angular resolution and sensitivity we needed, we used VLA B/C/D-array configurations
with 12/6/2 hr integration times per galaxy. 
The combined B/C/D arrays sample the galaxies at 6\arcsec\
(110 pc at 3.7 Mpc, the average distance of our sample; for {\sc robust}-weighting).
This resolution shows clouds, shells, and turbulent structures that
are important for star formation. The B array, which is the most extended configuration of the three we used,
requires the most time because of its reduced surface brightness sensitivity, 
but the high resolution it provides is essential.
Typical clouds and holes are of order 100 pc in diameter, and they
would be lost at the resolution of the data obtained in the more compact
C and D arrays (the resolution of C$+$D is $\sim$20$^{\prime\prime}$, or 360 pc at 3.7 Mpc).
However, the C and D arrays are important because they
reveal the extended, low-density gas around clouds, shells, and star formation.
Low spatial resolution maps also trace the low density \HI\ far beyond the stellar disks.
The D array can detect structure up to a scale of 15\arcmin.

For good velocity resolution, we chose a channel separation of 1.3 \kms\ and 2.6 \kms, both for
new observations and for data we drew from the archives.
The choice of channel separation was dictated by the need to contain galaxy emission
within the bandwidth, leaving line-free channels on either side for continuum subtraction.
Twenty-two galaxies have only Hanning-smoothed data cubes with 2.6 \kms\ channel separation, and
19 have small enough velocity ranges to be observed and imaged with 1.3 \kms\ channel separation.
These relatively small channel separations enable the resolution of kinematic structures
in the galaxies, especially the deconvolution of warm and cool components 
with velocity FWHM as small as 7-9 km s$^{-1}$(Young \et\ 2003).

In our sample of galaxies requiring new VLA observations, 
13 had no previous VLA observations, and 8 had some data in the archives.
Altogether, there were 21 galaxies that needed B array observations, 15 that needed C,
and 17 needed D configurations. 
Seven of our galaxies are part of the VLA-ANGST (ACS [Advanced Camera for Surveys] Nearby Galaxy Survey Treasury)
large project (Ott \et\ 2010; project ID AO215) 
that began at the same time as LITTLE THINGS: CVnIdwA, DDO 70, DDO 75, DDO 155, DDO 187, NGC 4163, UGC 8508.
The VLA-ANGST integration times were generally shallower than those of LITTLE THINGS, but
these galaxies were observed as part of VLA-ANGST rather than LITTLE THINGS and we use those data here.
In addition eight of our galaxies taken entirely from the archives were part of the THINGS (Walter \et\ 2008) sample 
(DDO 50, DDO 53, DDO 63, DDO 154, M81dwA, NGC 1569, NGC 2366, NGC 4214). 
For the THINGS galaxies, we retrieved their uv-calibrated data.

There are two galaxies that are exceptions to our combination of the B/C/D array configurations. 
These galaxies are DDO 216 and SagDIG. The data for these galaxies were taken from the 
archives where only C and D array configuration data were available.
        
\subsection{Ancillary data}

We also have a rich multi-wavelength collection of ancillary data on the LITTLE THINGS galaxies.
Our data sets are summarized in Table \ref{tab-datasets}.
We have two tracers of the products of recent star formation.
First, \ha\ images obtained at Lowell Observatory reveal the presence of massive stars formed
over the past 10 Myrs through the ionized nebulae of their natal clouds (Hunter \& Elmegreen 2004).
Second, ultraviolet images at 1516 \AA\ and 2267 \AA\ obtained with 
the {\it Galaxy Evolution Explorer} satellite ({\it GALEX}; Martin \et\ 2005) show 
the bright ultraviolet starlight from stars
formed over the past 100--200 Myrs (Hunter \et\ 2010).
The ultraviolet images are a more reliable tracer of recent star formation especially
in the outer disks of galaxies where star formation is more sporadic
and lower gas densities make \HII\ regions harder to detect.
The {\it GALEX} data were processed with the GR4/5 pipeline.

Older stars are traced with $UBV$ images obtained at Lowell Observatory (Hunter \& Elmegreen 2006). 
For on-going star formation, these are sensitive to stars formed over the past 1 Gyr.
For 17 of the galaxies, we also have some combination of $JHK$ images obtained at Lowell Observatory, 
and for 36 galaxies there are {\it Spitzer} 3.6 $\mu$m images that we have pulled from the archives.
The near and mid-IR images in particular
highlight the star formation activity integrated over a galaxy's lifetime.

Dust plays an important, although poorly understood, role in the star formation process.
FIR continuum emission traces this part of the ISM. 
{\it Spitzer} 5.8 $\mu$m and 8.0 $\mu$m images trace polycyclic aromatic hydrocarbon (PAH) 
emission (plus hot dust continuum emission).
{\it Spitzer} MIPS images of 35 of the galaxies,
obtained as part of the {\it Spitzer} Infrared Nearby Galaxies Survey (SINGS), 
Local Volume Legacy (LVL), and other programs,
can in principle give the dust continuum emission and distribution at 24 $\mu$m, 70 $\mu$m, and 160 $\mu$m,
although exposures tend to be shallow.
We are also mapping a subset of galaxies in the FIR with bolometer arrays on
the Atacama Pathfinder EXperiment (APEX) sub-millimeter telescope in collaboration with M.\ Rubio
and with {\it Herschel} (Pilbratt \et\ 2010; an OT2, priority 1 program) in collaboration with S.\ Madden.
These data are important for understanding the heating balance in the ISM.

Molecular clouds are a key component of the ISM and the component most directly connected with
star formation. We are exploring this part of the star formation process through case studies. 
This has included Combined Array for Research in Millimeter-wave Astronomy (CARMA) 
observations of CO line emission in NGC 3738, a dwarf with intense star formation in the
center of the galaxy. 
With APEX we mapped CO in WLM, 
a typical star-forming dwarf, in collaboration with M.\ Rubio.
We are also using {\it Herschel} (an OT1, priority 1 program) to map photo-dissociation regions (PDRs) in five of
the LITTLE THINGS galaxies in collaboration with M.\ Kaufman and S.\ Madden.
These data will allow us to characterize molecular clouds at low metallicity.

\section{VLA observations and data reduction}

\subsection{Observing}

The LITTLE THINGS (project AH927) observations began with 
the B-array 2007 December 19 and, for B-array, ended 2008 February 10.
The observations were done in two 6-hour segments for each galaxy. 
Some observations were repeated when technical hiccups occurred.
C-array observations followed in the spring of 2008,
and D-array observations in the summer of 2008.
All of our new observations were in hand by the beginning of August 2008.
The observations are listed in Table \ref{tab-obs}.
Each array configuration with which a galaxy was observed is listed separately. 
Multiple observing sessions for each configuration are identified by the dates
on which the observing sessions began.

Our new observations used a  bandwidth of 0.8 or 1.6 MHz (nominally 150 and 300 \kms\ wide), 
depending on the known velocity range of emission from each galaxy. 
We used two intermediate frequencies (IFs; 2AC) and 256 channels.

Our new data were not Hanning smoothed when the observations were made. 
Thus, the galaxies for which we 
were targeting a spectral resolution of 2.6 \kms\ (12.21 kHz) were observed with
a channel separation of 1.3 \kms\ (the spectral resolution is 1.4 times this), 
and in the reduction stage the data were Hanning-smoothed to obtain a channel separation and resolution of 2.6 \kms.
Similarly, galaxies for which we targeted a spectral resolution of 1.3 \kms\ (6.10 kHz)
were observed with a channel separation of 0.65 \kms\ (3.05 kHz) and later Hanning-smoothed
to obtain a channel separation and resolution of 1.3 \kms.
Most of the data from the archives were on-line Hanning smoothed.
The exception is SagDIG, which was observed with a channel separation of 3.05 kHz
and no Hanning-smoothing. For that galaxy, we produced a cube with the native resolution (1.4$\times$ the channel 
separation of 0.65 \kms),
and another cube with Hanning smoothing to produce a channel separation and resolution
of 1.3 \kms.

Because of the transition of the VLA to EVLA during the course of the survey, sensitivity changes, 
phase drifts, and phase jumps were expected. 
To deal with this required that careful calibration be built into the observations.
Our observations in general began and ended with a primary flux calibrator, usually 3C286, allowing extra time 
at the beginning in case of a cable wrap problem in moving to our galaxy position.
It was also recommended to add a short dummy
scan to make sure the antennas were properly initialized at the beginning of the observing session.
The primary calibrator was used to set the flux scale and for determining the response function across the bandpass.
In between these observations we alternated between the secondary calibrator and the galaxy
with approximately 12-25 minutes on source at any one time. The shorter on-source dwell times
were in the shorter observing sessions. 
Because of our narrow bandwidth, when offered with a choice, we preferred a strong calibrator,
even if it was at a radial distance that was larger than that of a more nearby, but weaker one.
This was possible because of the phase stability of the VLA at 20 cm even in the B-configuration.
Even so, the secondary calibrators were always
within about 12\arcdeg\ of a target galaxy. The secondary calibrator was used to determine the
variations of phase and amplitude with time during the observing session.
The time spent on source in each configuration is given in Table \ref{tab-obs}.

\subsubsection{The Aliasing Problem}

\begin{figure}
\epsscale{1.0}
\includegraphics[angle=0,width=1.0\textwidth]{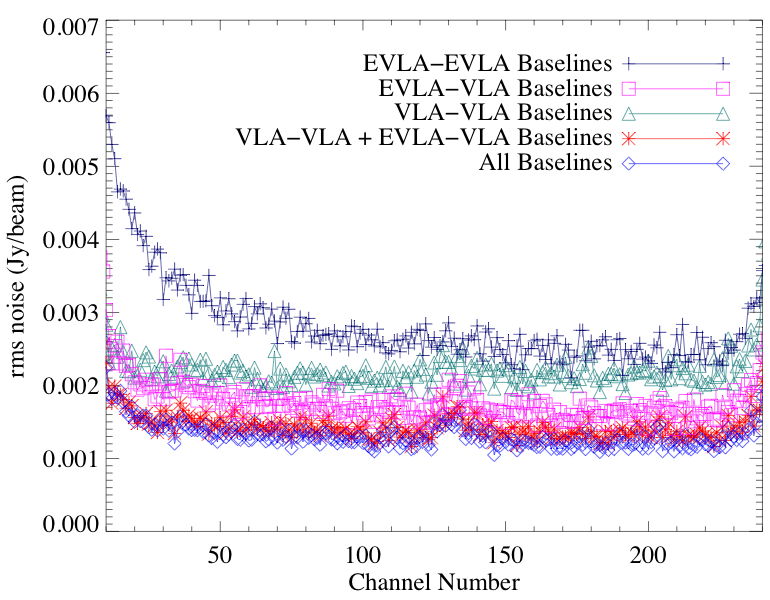}
\caption{Noise level as a function of channel number for data
obtained in B--configuration on DDO 43. The total bandwidth was
1.56 MHz split into 256 frequency channels. The observations were
made with 26 antennas, 12 of which were equipped with the new EVLA
front--ends, during the VLA--EVLA transition period. We show here
different combinations of baselines: combination of 66 EVLA--EVLA
baselines (navy crosses); 91 VLA--VLA baselines (green triangles); the
sum of 168 VLA--EVLA baselines (open magenta squares); the combination
of the VLA--VLA plus VLA--EVLA baselines (red stars); and finally all
325 baselines (open blue diamonds). 
Continuum emission was aliased, with phase information scrambled making it resemble
noise, the effect being strongest in channels closest to the baseband edge. This is most noticeable
on EVLA-EVLA baselines, and to a lesser extent on EVLA-VLA baselines; VLA-VLA baselines
were not affected.
\label{fig-noise}}
\end{figure}

During the course of our observations, the VLA was in transition to the EVLA,
and antennas were being outfitted with new receivers and electronics one by one.
At the beginning of our observations 46\% of the antennas in the array had been converted,
and by the end 62\% of the array consisted of EVLA antennas.
To make the digital signal from EVLA antennas compatible with the old analog VLA correlator, the signal
on the new receivers was passed through an electronic filter.
However, this conversion caused power from sidebands to be
aliased into the lower 700 kHz  of the observing bandwidth 
(the ``aliasing problem'' \footnote[15]{http://www.vla.nrao.edu/astro/guides/evlareturn/aliasing.shtml/}). 
In spectral line mode, the line emission would be unaffected
as long as the bandwidth used was
at least 1.5 MHz and the spectral line observed was narrow enough so that it fell beyond the affected 
0.7 MHz or the line could be shifted away from the band center. However,
even so, continuum emission was aliased, with phase information scrambled making it resemble
noise, the effect being strongest in channels closest to the baseband edge. This is most noticeable
on EVLA-EVLA baselines, and to a lesser extent on EVLA-VLA baselines; VLA-VLA baselines
were not affected (see Figure \ref{fig-noise}). 

This problem had a serious impact on LITTLE THINGS, which used narrow band-widths. 
A key element of our project is high spectral resolution
imaging at high sensitivity of these faint, low rotation velocity objects.
Thus, our galaxies
were being observed with small bandwidths (0.78 MHz or 1.56 MHz) in order to achieve 
high spectral resolution (1.3 \kms\ or 2.6 \kms).
We examined our data to understand the consequences of the aliasing problem and how
we might deal with it. We took 1.56 MHz bandwidth data
of one of our galaxies and fully calibrated and mapped it, running extensive tests.
We clearly saw increased noise from the
lower band edge to the middle of the passband when EVLA-EVLA and VLA-EVLA baselines
were included (Figure \ref{fig-noise}). 
In the end, we were able to satisfactorily calibrate the data only by eliminating
the EVLA-EVLA baselines. Subtracting the continuum from the line data using a
fourth order polynomial, as had originally been suggested, resulted in unacceptable amplitude
variations across the bandpass. An alternative scheme using unaffected baselines
to gauge the level of the continuum worked at some level, but continuum remained
in the map at unpredictable levels.
Furthermore, the noise added to the overall dataset by the EVLA-EVLA baselines
was at such an extent that the noise in a data set containing all baselines
and the noise in the same data set without EVLA-EVLA baselines are at virtually the same level. 
This means that, even if it were possible to calibrate properly
the EVLA-EVLA baselines, this would not reduce the noise level.
 
Because the EVLA-EVLA baselines constituted 22\% of the total in B-array,
we requested a 22\% increase in our B-array observing time,
or 55.5 hours, in order to make up for the lost EVLA-EVLA baselines and achieve our original target signal-to-noise.
This was granted in the form of dynamically scheduled time, in
blocks of 1--2 observing sessions (2.5--3 hr) per galaxy.
Similarly, time for C and D array observations were increased in proportion to the percentage
of EVLA-EVLA baselines lost, and this extra time was dynamically scheduled as well.
A consequence of the dynamic scheduling was that some of the make-up time came
during array configuration transitions and not when the instrument was in a ``pure'' configuration
(see Table \ref{tab-obs}).

We were worried that the loss of the EVLA-EVLA baselines might change the uv-coverage that we expected.
Comparing the uv-coverage with and without EVLA-EVLA baselines removed, 
however, showed that the antennas were being retrofitted in
a pattern that 
affected the uv-coverage only slightly and caused sidelobe structure that could be dealt with
in the imaging stage by employing a suitable CLEAN algorithm
(see Appendix B).

\subsubsection{Doppler Tracking}

Another consequence of the VLA-EVLA transition was that Doppler tracking was not possible.
Because the LO chains for EVLA and VLA front-ends are different, frequency changes can
cause phase jumps between EVLA and VLA antennas. Therefore, observing at a fixed frequency
was recommended.
This meant that the frequency of the observations changed by small amounts during the course of an 
observing session. We chose our central frequency for the expected date and time
of the observations, and checked to ensure that the galaxy would not move out of the channel
during the observations. Trickier were the dynamically scheduled observations. For those, we chose a
date in the middle of the range of possible observation dates. In no case did the galaxy emission move out of the channels,
but this added an additional concern in setting up the observe files and an additional step in mapping the data.

\subsubsection{Avoiding Milky Way Emission}

Another problem arose for six galaxies with 
radial velocities that are similar to those of \HI\ in the Milky Way in the direction
of the galaxy, potentially leading to Milky Way emission in the passband. 
These galaxies include DDO 165, DDO 167, DDO 187, NGC 4163, NGC 6822, and UGC 8508.
(NGC 1569 was observed before the transition to EVLA as part of THINGS.)
The standard approach in these cases is to observe both primary and secondary calibrators at velocities offset 
from that of Galactic \HI\ in their direction. This is accomplished by applying a frequency offset 
to a higher and lower central frequency, symmetrically with respect to the originally
intended central frequency. The offset needs to be large enough to shift the entire bandwidth
beyond the range where Galactic \HI\ is present. However, because phase jumps were expected to accompany
frequency shifts with the EVLA antennae, we observed only the primary calibrator with a positive and negative offset,
ensuring it was unaffected by Galactic foreground \HI. Because no absolute phase calibration is
required for the flux and bandpass calibration, the phase jumps accompanying any frequency shifts
were irrelevant for the primary calibrator.
For NGC 6822, there were two flux calibrators---3C286 and 3C48, and both were offset in frequency.
The offset was 3 MHz for 4 of the galaxies, and 3.5 MHz for DDO 187 and 2.5 MHz for NGC 6822.

Unfortunately, in calibrating these data we found a problem with
some of the calibrator observations on one side of the central frequency. From discussions with Ken Sowinski
at NRAO we realized that this was due to reassignment of the Local Oscillator (LO) settings by the on-line system 
that tunes the receivers, which meant the frequency range of the incoming signal fell partly beyond the range 
selected by the front end filter. We, therefore, had to throw out the corrupted observation of the calibrator and 
use only the frequency offset observation on the side which was not corrupted. 
This affects the flux calibration at a level of at most 0.1\% and can therefore be safely ignored.
     
\subsection{Calibration}

The VLA data were calibrated and mapped by many members of the team.
Usually one person would calibrate all of the data sets for a galaxy, and similarly one person
would usually complete the mapping of a galaxy.
As data sets were calibrated or the ensemble of data sets for a galaxy was mapped, 
the data were uploaded to an internet-accessible hard disk provided and maintained by NRAO, 
making it available to the rest of the team
and the next person who would work on the data. The final map cubes were then examined by
designated ``checkers'' who inspected all of the data, looking for signs of any problems or inconsistencies. 
A galaxy was not considered ``done'' until it had passed a thorough inspection, and any and all problems
were resolved. 

In order to ensure uniform treatment of the data, we created ``recipes'' for data
calibration and mapping that were carefully followed. Since the THINGS survey (Walter \et\ 2008)
had already been through this process and two of our team members were part of THINGS,
we started with their recipe and updated it to reflect changes in software,
complications due to the aliasing problem, and problems resulting from
the frequency shifts for observations that could be contaminated by Galactic \HI.
In order to ensure a homogeneous data set, we treated the archival data the same
way that we treated the new data.
All of the reductions were done in the Astronomical Image Processing System 
(AIPS) developed by NRAO, version 31DEC08.
We used the flux scale of Baars \et\ (1977).

We outline the salient calibration steps 
in Appendix A,
and the recipe itself is available on the public web site discussed in \S \ref{sec-public}.
After every action on the data, we had at least one, and usually multiple, quality control checks.
We outline the salient calibration steps, including those to check the quality of the calibration, in Appendix A. 
The recipe itself is available on the public web site discussed in \S 5.
For the cases involving multiple frequencies, most steps had to be run for each frequency.

A few circumstances yielded special calibration challenges:
1) Strong winds and many technical problems resulting in phase jumps meant that considerable data
had to be eliminated, limiting
the data obtained 2008 February 4, affecting DDO 101 (B array).
2) There was a clock error on data taken in B array 2008 January 5 and 7
and only VLA-VLA baselines could be used. The galaxies that were affected were
DDO 46, DDO 52, and DDO 133.
 A similar problem on 2008 January 24 rendered all data from EVLA antennas unusable, affecting DDO 87 and DDO 167.
 3) All of the EVLA antennas had to be flagged during the archive 2006 
 B array observations of VII Zw 403.  Four antennas were involved.
4) Solar interference affected the observations in compact configurations of some galaxies, 
usually those data $\leq$1 k$\lambda$:
DDO 43 (C and D arrays), DDO 47 (D array),
DDO 52 (D array), DDO 87 (C array), F564-V3 (D array), IC 1613 (C and D arrays),
NGC 3738 (D array), NGC 4163 (D array), WLM (B and C arrays), and VII Zw 403 (C and D arrays).
5) There were problems with the phases in the right polarization (RR), and some of the left (LL),
of the primary phase calibrator of DDO 43 (C array) in the archival data obtained in 2000 April.
To deal with this, we performed the bandpass calibration on the secondary instead of the primary calibrator,
flagged most of the left polarization data on the primary calibrator, 
kept only the left polarization in creating the new Channel 0 (an average of the inner 75\% of the
frequency channels),
ran the usual calibration on the primary and secondary calibrators to get the flux of the secondary calibrator,
ran the calibration {\sc calib} only on the secondary, and applied the calibration using {\sc clcal}
to the new Channel 0 data.
6) There was no model for the primary calibrator 3C295 used with VII Zw 403 in D array in the archival 1997 data.
However, the calibrator is not resolved in D array, so we used our observations of the calibrator instead of a model.
Similarly,
the secondary calibrators for archival observations of IC 1613 and of WLM in B array were spatially resolved. 
For IC 1613, we used the C array image of the calibrator
as a model for calibrating the B array with a maximum limit on the uv range of 15 k$\lambda$.
For WLM, the effect was small enough to ignore.
7) One galaxy, DDO 101, was self-calibrated 
in C and D arrays using a 2.3 Jy continuum source in the field.
8) The secondary calibrator, $0925+003$, of the archive data on DDO 70 had poor signal-to-noise (flux of $\sim$0.6 Jy).
9) For IC 10, the D array observation that we used is the central pointing ({\sc qual}$=$13) of a 25-pointing mosaic
(Wilcots \& Miller 1998).
     
\subsection{Imaging}

The mapping recipe picks up where the calibration leaves off, beginning with extracting the calibrated uv-data
for just the galaxy from the {\em uv}-data set ({\sc split}). 
The individual data sets taken in B, C, and D-configuration, were combined into one large {\em uv-}file, 
ready to be Fourier transformed to create maps in the image plane containing structure across a range 
of angular scales, from a few arcseconds up to several arcminutes. 
The spectra from different data sets were aligned in frequency space
({\sc cvel}), and the {\em uv}-data sets combined ({\sc dbcon}, with no adjustments to the relative weights). 
We subtracted continuum emission in the {\em uv}-plane ({\sc uvlsf}), followed by mapping.

The commonly used mapping/CLEAN algorithms for VLA data
approximate all emission structure as the sum of point sources. 
This is clearly an unrealistic approximation for extended galactic emission.  
In addition, the resulting map is the sum of the clean components that are made with the clean synthesized beam
and the residuals that are a product of the dirty beam. 
So, for mapping, we were excited by the advantages that multi-scale or multi-resolution CLEAN (MS-CLEAN) might bring.
This concept, developed by Wakker \& Schwarz (1988), solves simultaneously for point sources
and Gaussians with pre-determined sizes (Cornwell 2008).  This has been implemented
in AIPS under {\sc imagr} by Eric Greisen (Greisen \et\ 2009).
However, there has been limited experience with using this algorithm in the community (Rich \et\ 2008).
Therefore, before we could adopt it as our standard, we invested considerable time experimenting with the 
algorithm in order to understand the consequences to the data, in particular
comparing MS-CLEAN maps against those produced with standard CLEANing methods
(for more details see Appendix B).
We concluded that the MS-CLEAN algorithm gave superior maps and that we understood
it well enough to adopt it for making the LITTLE THINGS map cubes.
We then finalized our mapping recipe around the AIPS MS-CLEAN option in {\sc imagr}
(see details in Appendix B).

We produced both a Natural-weighted
image cube, which we expected would bring out extended lower surface brightness emission at the 
expense of angular resolution and beam profile,
and one weighted with a {\sc robust} value of 0.5 (Briggs 1995), which produces a 
better behaved synthesized beam at a resolution that is close to that obtained with Uniform-weighting.
Usually, the {\sc robust}-weighted cube was cleaned to 2$\times$rms and the Natural-weighted cube to 
2.5$\times$rms. These levels were chosen after extensive tests that showed that the flux that was recovered by MS-CLEAN
was not a strong function of CLEAN level, as long as that level was within factors of a few of the optimum level.
Cleaning deeper than 2$\times$rms gave noticeable CLEAN artifacts; cleaning down to 3-4$\times$rms  
left significant residual flux in the different scale maps. We converged on a level of 2$\times$rms.
However, the Natural-weighted cubes cleaned to 2$\times$rms showed signs of slight over-cleaning (``holes'' dug in the
background). For the Natural-weighted cubes we found that  2.5$\times$rms was the best compromise.
Each map cube was carefully inspected for signs of over- or under-cleaning. 

The maps of two galaxies, Mrk 178 and VII Zw 403, required small tweaks to this scheme. 
For Mrk 178, our standard CLEAN to 
2$\times$rms with the {\sc robust}-weighting produced 
a field at the 15\arcsec\ scale (see Appendix B) that appeared
to be over-cleaned. We, therefore, cleaned that scale to 2.5$\times$rms, but the other scales to 2$\times$rms.
For VII Zw 403, a CLEAN depth of 2$\times$rms resulted in being able to see the CLEAN box in the 
first two fields (0\arcsec, 15\arcsec), and so those limits were raised to 2.5$\times$rms. 
In both cases, the Naturally-weighted cube with the standard CLEANing depth of 2.5$\times$rms
produced good results.

For galaxies with all new data, we produced cubes at the observed spectral resolution and a second set
with Hanning-smoothing applied. The Hanning-smoothed version is presented here.
For cases where we were combining new data (not Hanning-smoothed) with archival data that were Hanning-smoothed,
we shifted the new data in frequency space with the Hanning-smoothing option turned on ({\sc cvel}).

To remove the portion of each channel map that was only noise, we used the Natural-weighted map cube 
to determine emission from the galaxy. We smoothed the original map cube to a beam of
25\arcsec$\times$25\arcsec\ ({\sc convl}) and blanked all pixels 
below $2$-$2.5$rms of the smoothed cube ({\sc blank}).
We then made a second pass, blanking by hand any additional emission that was clearly noise (for example,
features that do not appear in several consecutive spectral channels). This blanked cube served as the master
cube that was used as a conditional transfer for the cubes with smaller beam size.
We made flux-weighted moment maps from the resulting data cubes ({\sc xmom}): 
integrated \HI, intensity-weighted velocity field, and intensity-weighted velocity dispersion. 
We corrected the integrated \HI\ map for attenuation by the
primary beam ({\sc pbcor}) and replaced blanks with values of zero ({\sc remag}). 

A few galaxies were affected by Galactic emission. 
For VII Zw 403 Galactic emission is apparent in the C and D array data.
The channels containing Galactic emission but no VII Zw 403 emission were blanked.
However, Galactic emission in the channels that contain VII Zw 403 emission was not removed.
For NGC 1569 Galactic emission is a serious contaminant over a large velocity range. 
See Johnson \et\ (2012) for a discussion of how that galaxy was handled.

We had particular problems imaging DDO 75. There were 8 uv data sets including AB (when the configuration is
in transition from A to B), B, C, and D array configurations.
Six of the data sets were archival and two were obtained under project AO 215.
The moment maps have low level diagonal striping, and the integrated \HI\ maps have 
individual blanked pixels (``pinholes'') where there should be flux. 
The increased noise from the striping caused the algorithm that produced the moment maps ({\sc xmom})
to calculate unreasonable velocity values, and so those pixels were blanked in the moment map. 
We tried to identify the cause of the striping, but it appears at such a low level in the {\em uv}-data 
that we could not isolate and remove it. 
We concluded by cleaning to 3$\times$rms since cleaning to 2$\times$rms looked 
overcleaned and produced many more pinholes.

Information about the {\sc robust}-weighted and Natural-weighted map cubes are given in 
Table \ref{tab-obs}, along with the channel separation of the observations and the final number of frequency channels
in the cube. For each cube we give the beam full-width at half of maximum (FWHM) along the major and minor axes, the beam position angle,
and the rms in the maps. The rms in a single channel varies from 0.46 to 1.1 mJy beam$^{-1}$ in our sample.
The pixel size of the cubes is 1.5\arcsec, except
for DDO 216 and SagDIG, which were observed only in C and D array configurations and were mapped with 3.5\arcsec\ pixels
because of their significantly larger beam.
The map sizes are usually 1024$\times$1024 pixels.
A few particularly large galaxies on the sky or galaxies with nearby Galactic emission were imaged
at 2048$\times$2048 pixels.
         
\section{Basic Analysis} \label{sec-basicanal}

For each galaxy we have measured the total mass in \HI, determined the \HI\ flux-velocity profile, and 
measured the \HI\ surface density profile. These basic analysis products are presented here.
To make these measurements, we used the Groningen Image Processing System (GIPSY, Vogelaar \& Terlouw 2001).

The cube channel map units are in Jy beam$^{-1}$. This is related to brightness temperature in degrees Kelvin through
$$S({\rm Jy~ beam^{-1}}) = 1.65\times10^{-6} \Delta\alpha \Delta\delta ~ T_B (K)$$ (Brinks, private communication),
where $\Delta\alpha$ and $\Delta\delta$ are the major and minor axis FWHM in arcseconds given in Table 
\ref{tab-obs}.
The integrated \HI\ map is obtained after converting the channel maps to units of K km s$^{-1}$
using the above relation and summing along the velocity axis. The column density becomes
$$N_{HI}({\rm atoms ~ cm^{-2}}) = 1.823\times10^{18} \displaystyle\sum T({\rm K}) \Delta V ({\rm km\ s^{-1}}),$$ 
which becomes
$$N_{HI}({\rm atoms ~ cm^{-2}}) = 1.105\times10^{21} S({\rm Jy ~ beam^{-1} ~ m ~ s^{-1}}) / (\Delta\alpha \Delta\delta)$$
(Hibbard \et\ 2001).
The \HI\ mass profile is obtained by integrating the flux in each channel map over the area containing \HI\ emission 
and dividing by the number of pixels per beam area to arrive at S expressed in Jy.
The conversion to total mass is then
$$M_{HI}({\rm M_\odot}) = 235.6 ~ D({\rm Mpc})^2 ~ \displaystyle\sum S ({\rm Jy}) \Delta V ({\rm km\ s^{-1}}).$$
The factors for converting fluxes to masses for our {\sc robust}-weighted integrated \HI\ maps
are given in Table \ref{tab-masses}.

\subsection{HI masses}

To measure the total \HI\ mass, we summed the flux over the primary-beam corrected integrated \HI\ map made
from both the Natural-weighted image cube and the {\sc robust}-weighted cube ({\sc flux}).
Since the integrated \HI\ maps had been constructed from the blanked cubes and pixels without emission had values of zero,
the flux was summed over the entire image. The flux was then converted to mass.
The masses from the two maps agree well.
The average of the absolute value of the differences between the masses measured from the {\sc robust}-weighted maps 
and those measured from the Natural-weighted maps is 11\%,
with the {\sc robust}-weighted masses often higher than the Natural-weighted masses.

\begin{figure}
\epsscale{1.0}
\includegraphics[angle=0,width=0.85\textwidth]{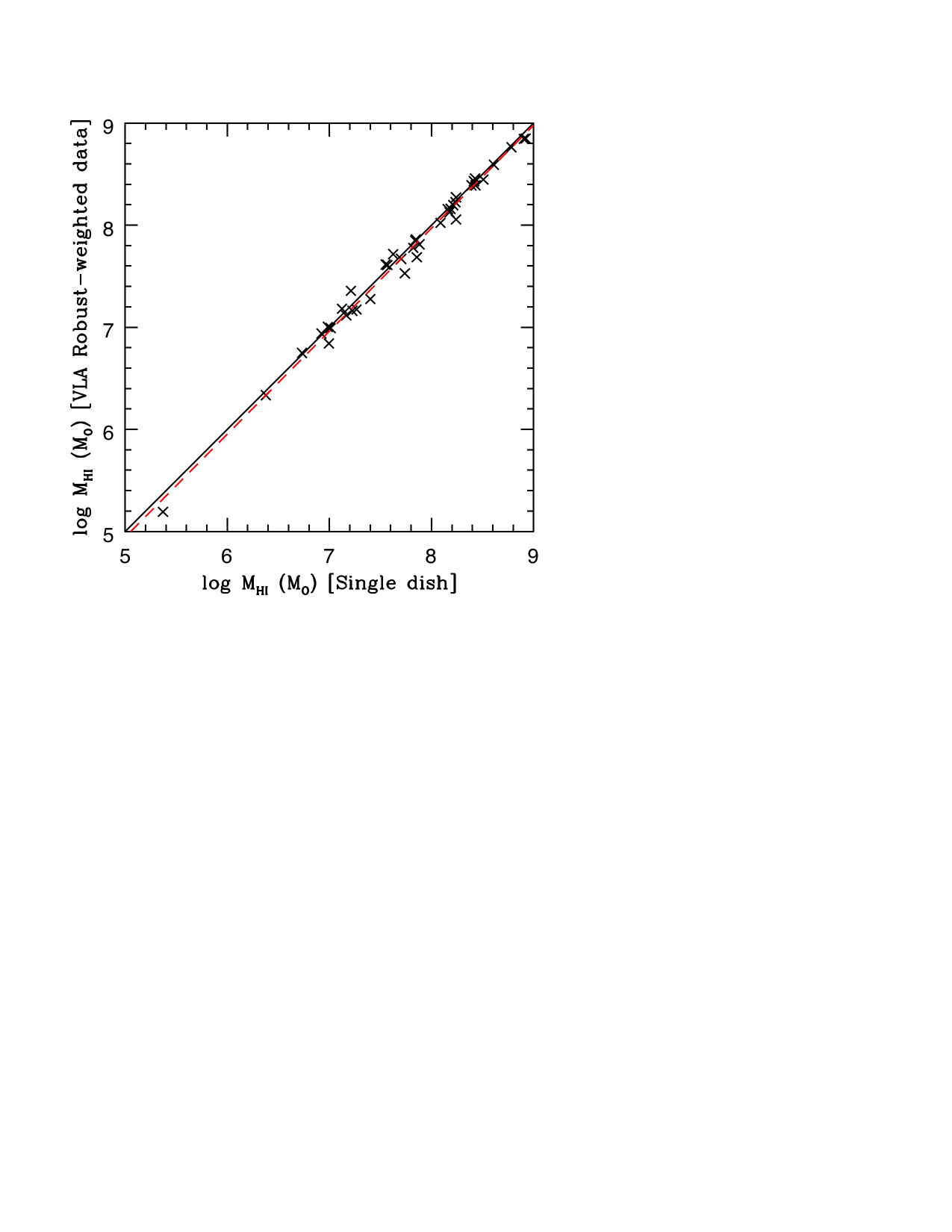}
\caption{Comparison of our integrated \HI\ masses with those from single-dish radio telescope observations.
Both measurements are relative to the same distance.
Our \HI\ masses are measured on the VLA integrated \HI\ {\sc robust}-weighted map.
The solid black line marks equal masses, and the red dashed line is the best fit to the data.
The average difference between the masses measured from the {\sc robust}-weighted maps 
and those measured from the Natural-weighted maps is 11\%. The best fit line lies
slightly below the line of equal masses, indicating that the VLA has systematically missed some flux compared
to single-dish values.
\label{fig-mass}}
\end{figure}

The fluxes and masses from the {\sc robust}-weighted data are given in Table \ref{tab-masses}, 
and compared to single-dish measurements in Figure \ref{fig-mass}.
The single-dish values are taken from Bottinelli \et\ (1982), Schneider \et\ (1990; DDO 101), 
Huchtmeier \et\ (2003; DDO 53, DDO 216, F564-V3), and Springob \et\ (2005; CVnIdwA).
Our VLA \HI\ masses agree within 11\% with those measured with single dish telescopes.
However, there is some bias towards VLA interferometer masses being slightly lower than those measured
with single-dish telescopes. 
One can see this in Figure \ref{fig-mass} by comparing the best fit line to the data in red with the
line of equal values in black.
     
\subsection{Velocity profiles and Single-dish comparison}

\begin{figure}
\epsscale{1.0}
\includegraphics[angle=0,width=1.0\textwidth]{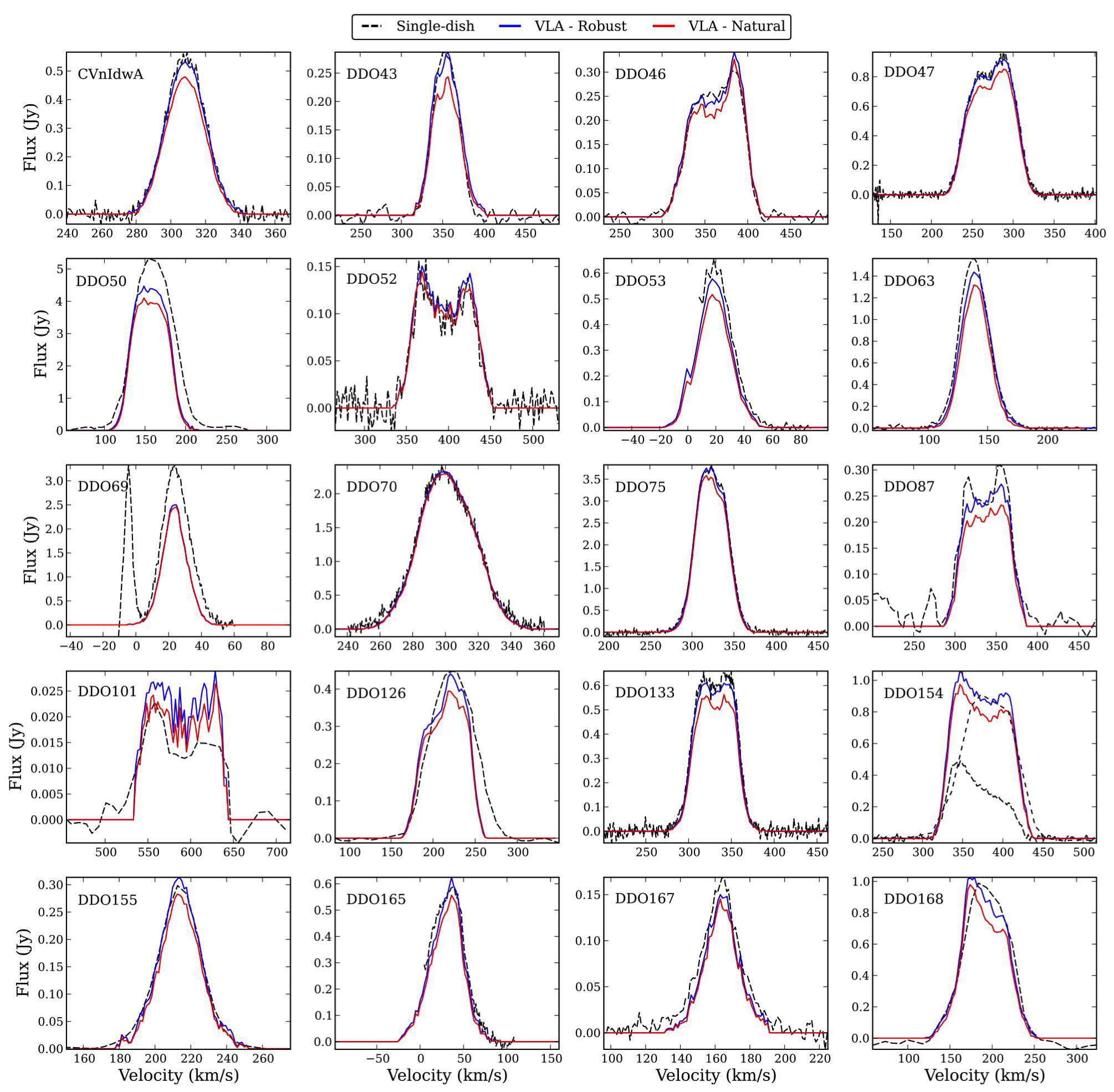}
\caption{Intensity-velocity plots for each galaxy made from summing the flux in each channel of the Hanning-smoothed
Naturally-weighted ({\it red}) and {\sc robust}-weighted ({\it blue}) \HI\ cubes. Single-dish profiles are also shown ({\it black}; 
see Table \ref{tab-velprofilecomp} for references).
\label{fig-velprof}}
\end{figure}

In general, we find good agreement between the VLA and single-dish measurements in the literature.  
We used spectra from the NASA Extragalactic Database, when available, and 
the telescopes that the single-dish measurements come from are given in Table \ref{tab-velprofilecomp}.
When comparing velocity profiles, we use our VLA {\sc robust} maps as they generally recover more flux than the 
Natural-weighted maps, which were cleaned to 2.5$\times$rms rather than 2$\times$rms.  
The profiles are shown in Figure \ref{fig-velprof} and details of the comparison of individual galaxies
are listed in Table~\ref{tab-velprofilecomp}.

Galaxies that show no significant difference between the interferometer and single-aperture measurements include 
CVnI dwA, DDO 43, DDO 46, DDO 47, DDO 52, DDO 70, DDO 75, DDO 133, DDO 155, DDO 210, DDO 216, SagDIG, and VII Zw 403.  
Those that are very close in shape, but have 10\% to 25\% less emission in the VLA profiles include DDO 53, DDO 63, DDO 87, DDO 167, and DDO 187.  
Most of these differences fall within the range of measurement uncertainty, and 
the choices of baseline fits can also account for some of the discrepancies.  
Some galaxies (LGS3, M81dwA, NGC 3738, UGC 8508, Haro 36, Mrk 178) have rather noisy profiles,
but generally agree in shape.

Galactic \HI\ emission affects several of the single-dish profiles.  
For DDO 53, DDO 69, Sag DIG, and VII Zw 403, the addition is obvious and 
far enough away in velocity from the peak of the extragalactic emission that it introduces no confusion in the comparison.  
NGC 1569, however, is heavily affected by Galactic \HI, so that the single-dish and interferometric profiles are hard to compare.

IC 10 and IC 1613 exhibit much more emission in the single-dish observations, with the VLA only recovering 40\% to 50\% of the
peak single-dish flux.  The VLA could be missing a significant amount of emission simply because the galaxies have large angular extents.  
Interferometers are inherently less sensitive to emission filling a substantial fraction of the primary beam.
DDO 50, DDO 69, DDO 126, DDO 165, NGC 4163, and NGC 4214 are other galaxies that 
show more modest examples of this behavior.  The high-velocity gas in Haro 29 matches quite well with single-dish observations, 
but at the low-velocity side of the profile, the single-dish observations have a broad wing of low brightness material that does not 
show up in the VLA profile.

For galaxies that occupy a larger area on the sky than the half-power beam width of a single-dish instrument, we
expect the VLA, with its relatively large HPBW, to recover more extended flux.  This can show up in the form of 
enhanced ``shoulders'' or asymmetric excesses in the VLA data.  
Galaxies in this situation are DDO 101, DDO 154, DDO 168, F564-V3, NGC 2366, and WLM.  
Two single-dish profiles are presented for DDO 154 because there is some disagreement between the 
published single-dish data.  
          
\subsection{Surface mass densities}

\begin{figure}
\epsscale{1.0}
\includegraphics[angle=0,width=0.9\textwidth]{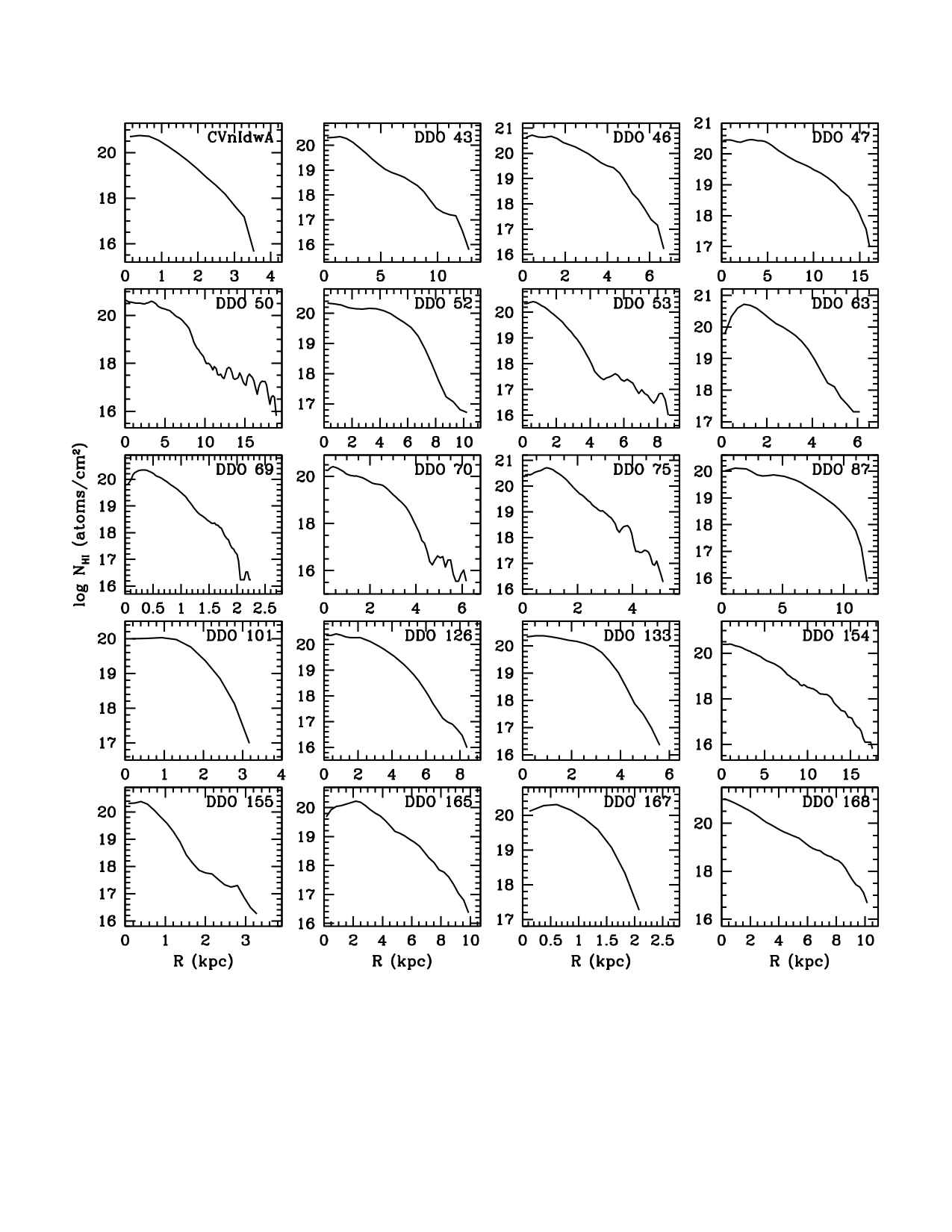}
\caption{Azimuthally-averaged \HI\ surface density as a function of distance from the center of the galaxy.
These are measured from the {\sc robust}-weighted velocity-integrated \HI\ maps, 
using parameters---center position, position angle of the major axis, and inclination of the
galaxy---determined from the optical $V$-band morphology (see Table \ref{tab-sd}; Hunter \& Elmegreen 2006). 
The width of the annuli and step size for the radius of each annulus was chosen to be approximately 
that of the FWHM of the VLA beam.
\label{fig-densprof}}
\end{figure}

To construct the \HI\ mass surface density profiles, we measured the flux in 
annuli on the velocity-integrated \HI\ maps ({\sc ellint}). 
We did this using the optically-defined parameters of center position, position angle of the major axis, and inclination of the
galaxy. These parameters were determined from the optical $V$-band morphology (Hunter \& Elmegreen 2006),
and are given in Table \ref{tab-sd}.
\HI\ kinematic parameters are in the process of being measured as the velocity fields are determined (Oh \et, in preparation),
and eventually surface density profiles will be made with the kinematic parameters. 
The optical inclination assumes an intrinsic minor-to-major axis ratio $b/a$ of 0.3
(Hodge \& Hitchcock 1966, van den Bergh 1988).
The width of the annuli and step size for the radius of each annulus
for each galaxy was chosen to be approximately that of the FWHM of the
beam, and the profiles were integrated from the center of the galaxy outward until we ran out of image.
The surface density profiles for the {\sc robust}-weighted integrated \HI\ maps are shown in Figure \ref{fig-densprof}.
         
\section{Public data products} \label{sec-public}

The \HI\ and ancillary data are being made available to the public through NRAO.
The \HI\ data include the Naturally-weighted and {\sc robust}-weighted cubes and
moment maps (velocity-integrated \HI\ maps \HI, intensity-weighted velocity field, intensity-weighted velocity dispersion).
The ancillary data include FUV, NUV, $UBVJHK$, and \ha\ images, as available for each galaxy.
In addition to the images and cubes, we will include basic analysis products in tabular and graphical form as they become
available. These currently include the
velocity-integrated flux profiles and \HI\ mass surface density profiles.

The NRAO web site is http://science.nrao.edu/science/surveys/littlethings, 
and the data can be obtained there now.
The ancillary data are also available at the Lowell Observatory web site http://www.lowell.edu/users/dah/littlethings/index.html.
We believe these data will be useful to the community for a wide variety of investigations.

\acknowledgments
We thank all of the people at NRAO and specifically the staff at the VLA who worked hard to make the LITTLE
THINGS VLA observations possible, and for the support we have received since the end of the
observations in dealing with various problems, setting up a team data site, and now supporting
a web site for public access of the data.
This work was funded in part by the National Science Foundation through grants AST-0707563, 
AST-0707426, AST-0707468, and AST-0707835 to DAH, BGE, CES, and LMY.
The {\it GALEX} work was funded by 
NASA through grant NNX07AJ36G and by cost-sharing from Lowell Observatory.
VH was supported by a Post-Doctoral Research Assistantship from the UK Science and Technology Facilities Council (STFC).
Parts of this research were conducted through the Australian Research Council Centre of 
Excellence for All-sky Astrophysics (CAASTRO), through project number CE110001020.
We are grateful to Ms.\ Lauren Hill for making the false color composite pictures in Figures 12-90.
This research has made use of the NASA/IPAC Extragalactic Database (NED) which is operated by the 
Jet Propulsion Laboratory, California Institute of Technology, under contract with the 
National Aeronautics and Space Administration.

Facilities: \facility{VLA} \facility{Hall} \facility{Perkins} \facility{GALEX} \facility{Spitzer}

\appendix \label{appendix-details}

\section{Calibration and Combination of Data}

The calibration steps followed broadly standard procedures; 
here we highlight those steps that needed particular attention and our ``recipes'' are available online.
In what follows, CH0 refers to an average of the inner 75\% of the frequency channels observed; 
LINE refers to the spectral line data set.

Data were loaded into AIPS using {\sc fillm} with the option to store channel/IF dependent weights with the visibilities, 
creating a CH0 and LINE file. The CH0 data were inspected for interference and other obvious problems using {\sc tvflg}. 
The visibilities were updated with baseline corrections from {\sc vlant}. The EVLA-EVLA baselines were flagged in their entirety.

We removed 8\% of the channels, 4\% on either side of the band, from the LINE data using {\sc uvcop} because the
noise in the first channel would percolate through to the bandpass calibration.
We subsequently proceeded with the bandpass calibration. Contrary to what was customary when reducing VLA data, 
we did the bandpass correction as one of the first steps, rather than as one of the last.
This was to remove VLA-EVLA closure errors due to non-matched bandpass shapes between VLA and EVLA antennas.

We recreated a new CH0 using {\sc avspc} on the LINE data and applying the bandpass correction. 
It is this CH0 that we then further calibrate in a fairly standard manner. 
We ran a number of diagnostic tests to check the quality of the calibration and to detect  possible problems, 
notably those related to the VLA-EVLA transition period 
Quality control checks 
included using {\sc possm} to examine the band-pass calibration done prior to constructing a new Channel 0 (CH0)
(to examine the band-pass calibration table, to apply the calibration to the secondary calibrator and 
examine individual baselines, and to apply to the secondary calibrator as a vector average of all data),
{\sc snplt} to inspect the amplitude and phase calibration (SN) table,
{\sc listr} to look at the SN table itself and check for phase or amplitude jumps, 
{\sc anbpl} to check the antenna-based data weights,
{\sc tvflg} to examine the uv-data as a function of baseline and time,
{\sc uvplt} and {\sc wiper} to plot the amplitude and phase against uv-distance for the calibrators and galaxy with the CH0 and
with the line data,
{\sc imagr} to map the calibrators and check their fluxes and map the source with the CH0 and line data,
{\sc possm} to check the vector average of all of the data for each calibrator,
and computing the actual and expected S/N.

With few exceptions (see Table \ref{tab-obs}) the LITTLE THINGS galaxies were observed in three different 
array configurations, B, C, and D, that needed to be combined into one data set. 
Furthermore, multiple observing runs contributed to the data taken in each of the configurations. 
Some of the data came from the NRAO archive, whereas the rest were newly observed. 
The archival data were Hanning-smoothed whereas the new data were not.

Once calibrated, we ran task {\sc cvel} to correct the frequencies from scan to scan as well as to align all observations 
of each galaxy in velocity. If newly observed data were to be combined with archival observations, 
Hanning-smoothing was applied to the data as well, again as part of task {\sc cvel}. 
Although this task applies  Hanning-smoothing, it does not remove every other channel, which is what used to 
be the case with the VLA correlator when specifying the Hanning-smoothing option. 
To accomplish this we used task {\sc uvdec}. Once the data sets had the same frequency resolution and sampling, 
they were aligned in velocity and, where required, precessed to the same J2000 epoch using {\sc uvfix}. 
We used task {\sc dbcon} to merge them into the final dataset.

We used {\sc uvlsf}, and a linear baseline, to extract the continuum based on the location of channels free from line 
emission as identified in a dirty image cube of all the combined data.

\section{Imaging}

Interferometers measure the Fourier transform of the sky intensity distribution. 
However, interferometric data are affected by incomplete coverage of the aperture plane: missing zero-spacing flux and 
information on short-baselines, missing or deleted baselines, missing or deleted hour angle ranges. 
To remove the effects this incomplete coverage has on the resulting images, deconvolution algorithms were developed, 
notably  the method known as CLEAN (H\"ogbom 1974). Although powerful, CLEAN  has its limitations, 
some of which we mention briefly here before we describe the approach we took.

\begin{figure}
\epsscale{1.0}
\includegraphics[angle=0,width=1.0\textwidth]{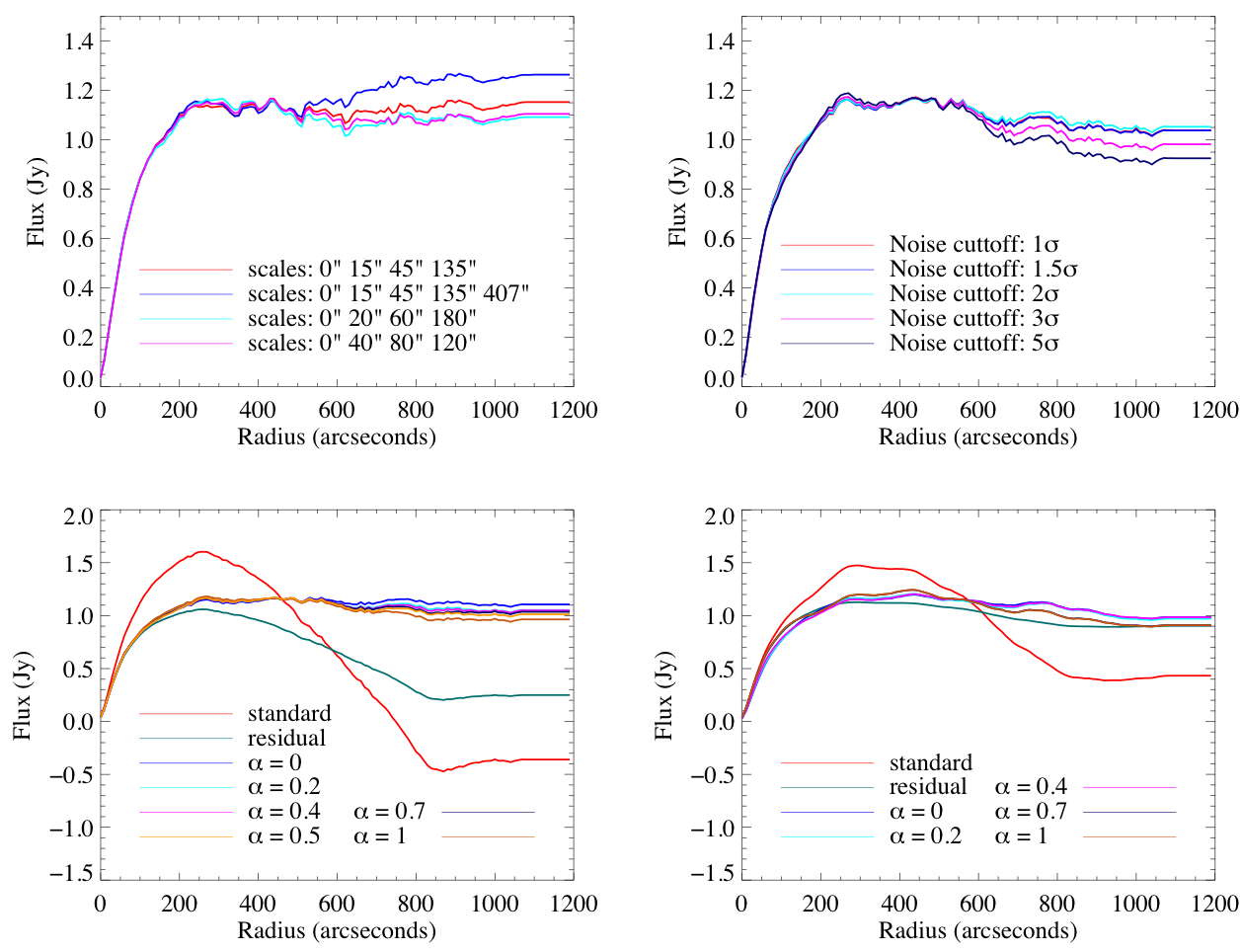}
\caption{Integrated Flux density  versus radius in DDO 168, channel
60, B+C+D--configurations combined. {\em Top Left:} Results obtained
using MS-CLEAN with $\alpha =0.2$, robust = 0, and testing different
combinations of angular scales 
(red: 0, 15\arcsec, 45\arcsec, 135\arcsec; 
blue: 0, 15\arcsec, 45\arcsec,
135\arcsec, 407\arcsec; cyan: 0, 20\arcsec, 60\arcsec, 180\arcsec; 
magenta: 0, 40\arcsec, 80\arcsec, 120\arcsec). 
{\em Top
Right:} Results obtained using MS-CLEAN with $\alpha=0.2$, robust =
0, and testing variable flux cutoffs (red: $1\sigma$;  blue:
$1.5\sigma$; cyan: $2\sigma$; magenta: $3\sigma$; navy: $5\sigma$).
{\em Bottom Left:} Results obtained by cleaning down to $2\sigma$,
robust = 0, and testing dependence on $\alpha$ (red: standard CLEAN,
green: residual scaled CLEAN, blue: MS-CLEAN and $\alpha = 0$, cyan:
$\alpha = 0.2$, magenta: $\alpha = 0.4$, orange: $\alpha = 0.5$, navy:
$\alpha = 0.7$,  brown: $\alpha = 1$). {\em Bottom Right:} As Bottom
Left, but with natural weight.
\label{figC1}}
\end{figure}

Standard CLEAN decomposes an image into clean components (basically Dirac delta functions at grid point locations) 
and a residual map. The clean components convolved with the original (also known as dirty) beam, 
a process known as restoring, and put back on to the residual map would give the original (dirty) image back. 
If instead the restoration is done with a Gaussian beam that matches the resolution of the dirty beam, a cleaned map results. 
Usually the flux in the cleaned map is determined under the assumption that the clean beam applies,
which is only true if all flux registered by the interferometer was cleaned. 
Because noise in the map makes it often difficult to clean down to the noise level, the 
residual map can contain appreciable residual signal, especially when cleaning
extended structures (J\"ors\"ater \& van Moorsel 1995). The flux is calculated correctly only for the clean components 
retrieved from the map, for the residual map it is overestimated by a factor equal to the ratio between the dirty and 
clean beam (for more details, see Walter \et\ 2008).  For this reason, a standard clean using AIPS task {\sc imagr} 
can overestimate the flux by as much as 50\% as shown in Fig.~\ref{figC1} (bottom left panel, red line). 
Rescaling the flux in the residual map can be used to correct for this (Fig.~\ref{figC1}, bottom left panel, green line), 
but does not deal with the fact that a substantial amount of flux has not been retrieved, and hence cleaned. 
This manifests itself by the emission sitting within a depression in the map (commonly referred to as the negative bowl), 
and in Fig.~\ref{figC1} (bottom left panel), by the subsequent drop after reaching a maximum when integrating a 
map out to ever larger radii.

Extended, low--level emission is a major contributor to the integrated flux of extended sources, even if the brightness in each pixel 
is at low signal--to--noise, which means that CLEAN would have to go to very low levels to recover all signal. That is  difficult as 
CLEAN tends to diverge when reaching levels below typically $2\times$rms. To address this, Wakker \& Schwarz (1988) 
developed an algorithm that solves sequentially, in the image plane, for structures at increasing scale. 
Their algorithm formalized the concept introduced by Brinks \& Shane (1984) under the name of 
Multi--resolution CLEAN, who ran CLEAN on an image at full resolution down to a set limit, after which the residual image 
and dirty beam were convolved to lower resolution and cleaned further, and so on, cleaning at 3 or 4 resolutions in total.

An extension of the Wakker \& Schwarz (1988) algorithm was further developed by Cornwell as Multi--Scale CLEAN 
(Cornwell \et\ 1999;  Cornwell 2008) and implemented in AIPS by Greisen as a Multi--resolution CLEAN within 
AIPS task {\sc imagr} (Greisen \et\ 2009). An early implementation within the Common Astronomy Software Applications 
(CASA) package was tested on galaxies that are part of THINGS (Rich et al. 2008). 
We will refer to both of these implementations as Multi--Scale CLEAN (or MS-CLEAN for short).

MS-CLEAN deals with extended structure by using {\em a priori} knowledge of spatial scales expected to be present in the image. 
When smoothing, the ratio of the extended source structure to beam size improves, 
which means that what was extended in the original resolution 
becomes small scale in a smoothed map. 
Moreover, the signal to noise in a smoothed map improves as long as the 
structure in the map remains resolved, allowing a much deeper clean and therefore ensuring that most flux is recovered. For
information on all scales to be retrieved in the AIPS implementation of MS-CLEAN, the user not only can choose the widths of 
circular Gaussians that will be used to make beam and data images at each scale, but also the relative importance given to small 
versus large scales.

The AIPS implementation of MS-CLEAN uses the multi--facets approach available in {\sc imagr} for wide--field imaging. 
Whereas for wide--field imaging {\sc imagr} creates a map made of adjacent facets, in MS-CLEAN mode {\sc imagr} creates a 
map out of overlapping facets over the same area of the sky, each facet representing a different resolution scale instead. At each 
major cleaning cycle, it selects one of the angular scales, re--images that facet with the current residual $uv-$data, finds clean 
components for that facet using minor cycles and continues until it reaches one of the stopping criteria, prompting clean to go 
through the next major cycle. The selection of which resolution should be cleaned is done by establishing which resolution has the 
highest peak flux. The algorithm stops when the flux level in all facets (at each of the resolution scales) is below the user--defined 
flux cut--offs or the number of components cleaned has reached a set limit. Then, the clean components for all resolutions, each 
with their corresponding clean beam, are restored to the image made at the highest resolution (Greisen et al. 2009).

It should be stressed that the parameter choice should be carefully tuned to the data set to be imaged. The algorithm  available 
under the {\sc imagr} task in AIPS requires a number of input parameters. The first two of them are {\sc ngaus} and {\sc wgaus} 
representing the number of scales to clean simultaneously and the approximate width of the convolution kernel used, in 
arcseconds. We applied the algorithm to images with data collected in B, C, and D-configurations, i.e., with a substantial range of 
angular scales (spatial frequencies). After several tests (see Fig.~\ref{figC1}, top left panel) we decided to use four kernels with 
widths of 0$\arcsec$ (i.e., native resolution), 15$\arcsec$, 45$\arcsec$, and 135$\arcsec$.

{\sc fgaus} allows the user to impose a stopping criterion for each resolution. We established our rms noise threshold by measuring 
the noise at each of the four resolutions of a dirty, line--free channel. Consequently, we have four different noise levels, one for 
each field. The four cut--off levels to be set in FGAUS are not independent, being different smoothed versions of the same dataset. 
We investigated different flux cut--offs, trying both deeper cleaning, down to 1$\times$rms and 1.5$\times$rms and more superficial 
cleaning to 3$\times$rms and 5$\times$rms. We plot the integrated flux as a function of radius in Fig.~\ref{figC1} (top right panel). We find 
that cleaning down to 3 or 5$\times$rms is not sufficient as we are not retrieving all the flux and evidence for a ``negative bowl'' is seen 
at large radii. Cleaning down to 1 or 1.5$\times$rms does not retrieve more flux than cleaning down to 2$\times$rms; however, clean 
artifacts start to become noticeable in the final image. This implies we do not need to go any deeper than 2$\times$rms 
to retrieve all the observed flux. 
We therefore decided for LITTLE THINGS on cut--off levels of 2$\times$rms for {\sc robust}-weighted data cubes and 2.5$
\sigma$ for the Natural-weighted cubes.

Naively one would always look for the peak residual at any resolution, and choose the next clean component based on the highest 
value found. However, since the lower resolutions dominate in terms of flux, the algorithm would be biased towards the largest 
scales (Greisen et al. 2009). 
To balance this, peak fluxes can be weighted by a factor of 1/[(field beam area)/(minimum beam area)]$^
\alpha$. Here $\alpha$ provides control over which resolution is going to be chosen next. If $\alpha=0$, the peak fluxes at 
each resolution are unaltered; if $\alpha=1$, the peak flux at the highest resolution will be the true peak flux, while the peak flux at 
all other resolutions will be down weighted. In the AIPS implementation of MS-CLEAN, $\alpha$ is set by the parameter {\sc 
imagrprm(11)}.

\begin{figure}
\epsscale{1.0}
\includegraphics[angle=0,width=1.0\textwidth]{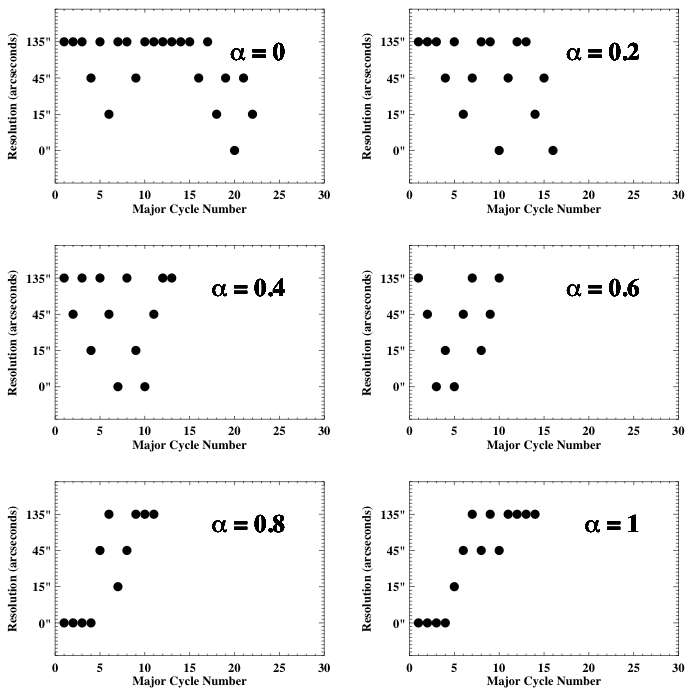}
\caption{Testing the $\alpha$ parameter on DDO 168, channel 60,
B+C+D--configurations combined. 
We plot the choice of resolution scale in each major cycle for $\alpha=0, 0.2, 0.4, 0.6, 0.8, 1.0$.
The parameter $\alpha$ provides control over which resolution is chosen next. 
If $\alpha=0$, the peak fluxes at each resolution are unaltered; 
if $\alpha=1$, the peak flux at the highest resolution will be the true peak flux, while the peak flux at 
all other resolutions will be down weighted. 
We found that a value of $\alpha=0.2$ strikes the right compromise, 
with the clean alternating among the different spatial scales and thus finding emission across a range of  angular scales 
rather than being fixated on one particular value. 
\label{figC5}}
\end{figure}

To illustrate how this biasing works we plot the choice of resolution in each major cycle when cleaning one channel 
(Fig.~\ref{figC5}). We ran tests on the choice of $\alpha$ and we found that for all $\alpha$, the cleaning process converges. 
Also, the total retrieved flux is very similar. We found that a value of $\alpha=0.2$ strikes the right compromise, 
with the clean alternating among the different spatial scales and thus finding emission across a range of  angular scales 
rather than being fixated on one particular value. The effect of different values for $\alpha$ on the integrated flux is shown in 
Fig.~\ref{figC1} for both {\sc robust} and Natural-weighted data.

\begin{figure}
\epsscale{1.0}
\includegraphics[angle=0,width=1.0\textwidth]{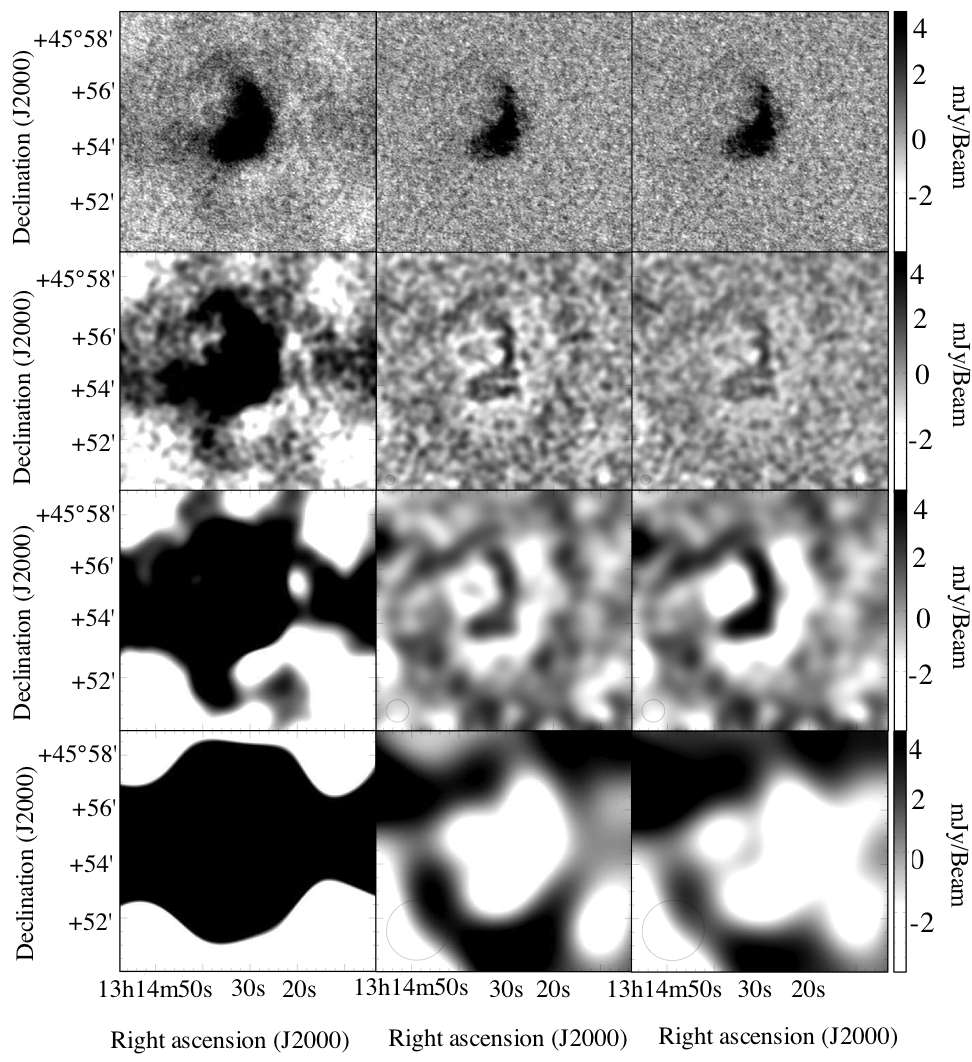}
\caption{DDO 168, channel 60, B+C+D--configurations combined:
Comparison between 2$\times$rms cut--off level, robust 0.5, clean maps
({\it top}) and residual maps for $15^{\prime\prime}$ ({\it second row}),
$45^{\prime\prime}$ ({\it third row}) and $135^{\prime\prime}$ ({\it fourth row})
for $\alpha=0.2$ ({\it middle column}) and $\alpha=0.4$ ({\it right column}). The
beam size is shown in the bottom left corner of each figure of the
middle and right columns. In the first column we show as reference the
uncleaned map for each field. All panels of the figure use the same
grey scale, shown on the right hand side.
\label{figC2}}
\end{figure}

\begin{figure}
\epsscale{1.0}
\includegraphics[angle=0,width=1.0\textwidth]{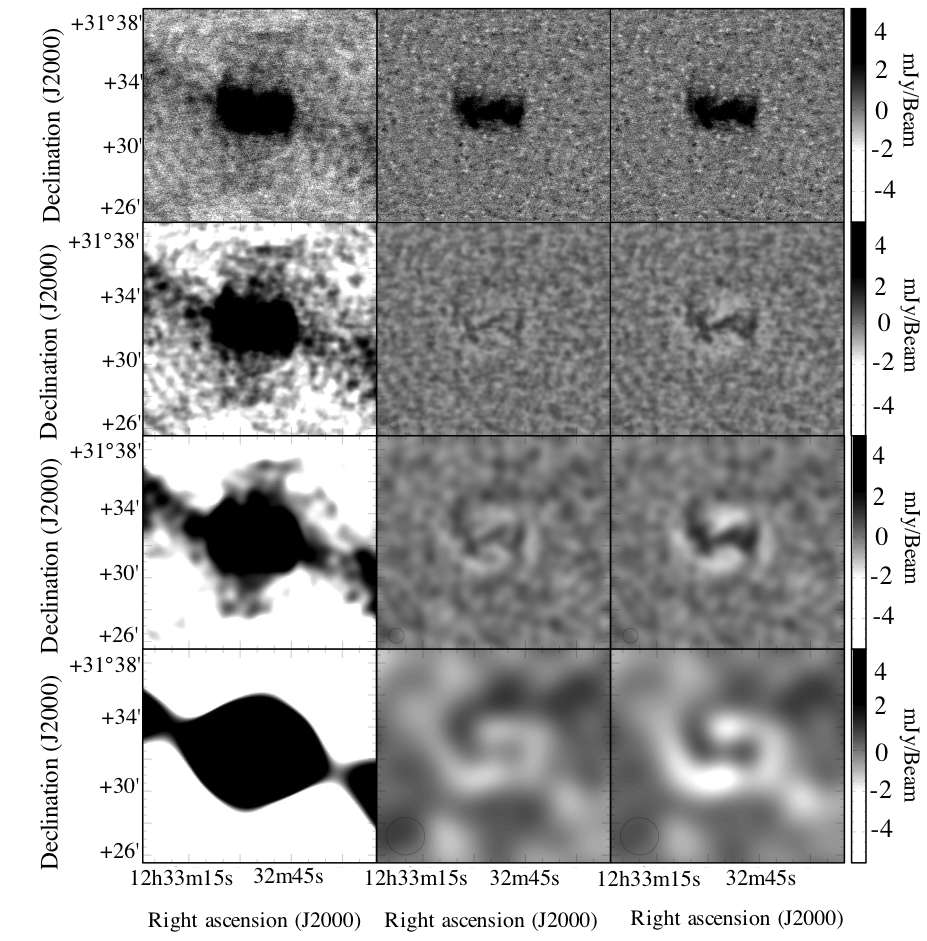}
\caption{DDO 133 channel 56, B+C+D--configurations combined:
Comparison between 2$\times$rms cut--off level, robust 0.5, clean maps
({\it top}) and residual maps for $15^{\prime\prime}$ ({\it second row}),
$45^{\prime\prime}$ ({\it third row}) and $135^{\prime\prime}$ ({\it fourth row})
for $\alpha=0.2$ ({\it middle column}) and $\alpha=0.4$ ({\it right column}). The
beam size is shown in the bottom left corner of each figure of the
middle and right columns. In the left column we show as reference the
uncleaned map for each field. All panels of the figure use the same
grey scale, shown on the right hand side.
\label{figC4}}
\end{figure}

The differences between $\alpha=0.2$ or 0.4 become more noticeable when looking at the final and residual maps resulting from 
applying MS-CLEAN to a line channel for two different sources, DDO 168 and DDO 133 (see Figs.~\ref{figC2} and~\ref{figC4}). 
It is clear that a particular choice for $\alpha$  depends on the complexity of the source to be imaged. Although there is 
little difference in the final maps for DDO 168, in the case of DDO 133 there are signs of over--cleaning for $\alpha=0.4$.

As an aside, it is interesting to note to what extent the residual scaled standard CLEAN follows  
the Natural-weighted MS-CLEAN results, even at larger 
radii, which was not the case for the {\sc robust}=0 MS-CLEAN case (see Fig.~\ref{figC1}, bottom right panel). 
This is not an effect of MS-CLEAN, but 
rather a consequence of the different density of visibilities near the center of the $uv-$plane and the different manners
this is dealt with
in the weighting scheme employed, Natural versus {\sc robust}.

Other than those described above, all further parameters were left at their default values. If we compare MS-CLEAN and traditional 
CLEANing methods we find that in our test the standard CLEAN has a peak flux which is 40\% higher than the MS-CLEAN or  
residual scaled CLEAN (see Fig.~\ref{figC1}, bottom left panel). This shows that MS-CLEAN, unlike standard CLEAN, properly 
addresses the problem related to determining the total flux. At the same time MS-CLEAN recovers all measured flux much better 
than the standard CLEAN, reducing considerably the negative bowl problem in extended sources. As a caveat, MS-CLEAN does 
not recover flux that was not measured at spacings shorter than the shortest spacing partaking in the observation. Therefore, any 
emission on scales more extended than those present in the observation will be missing.

\begin{figure}
\epsscale{0.85}
\includegraphics[angle=0,width=0.8\textwidth]{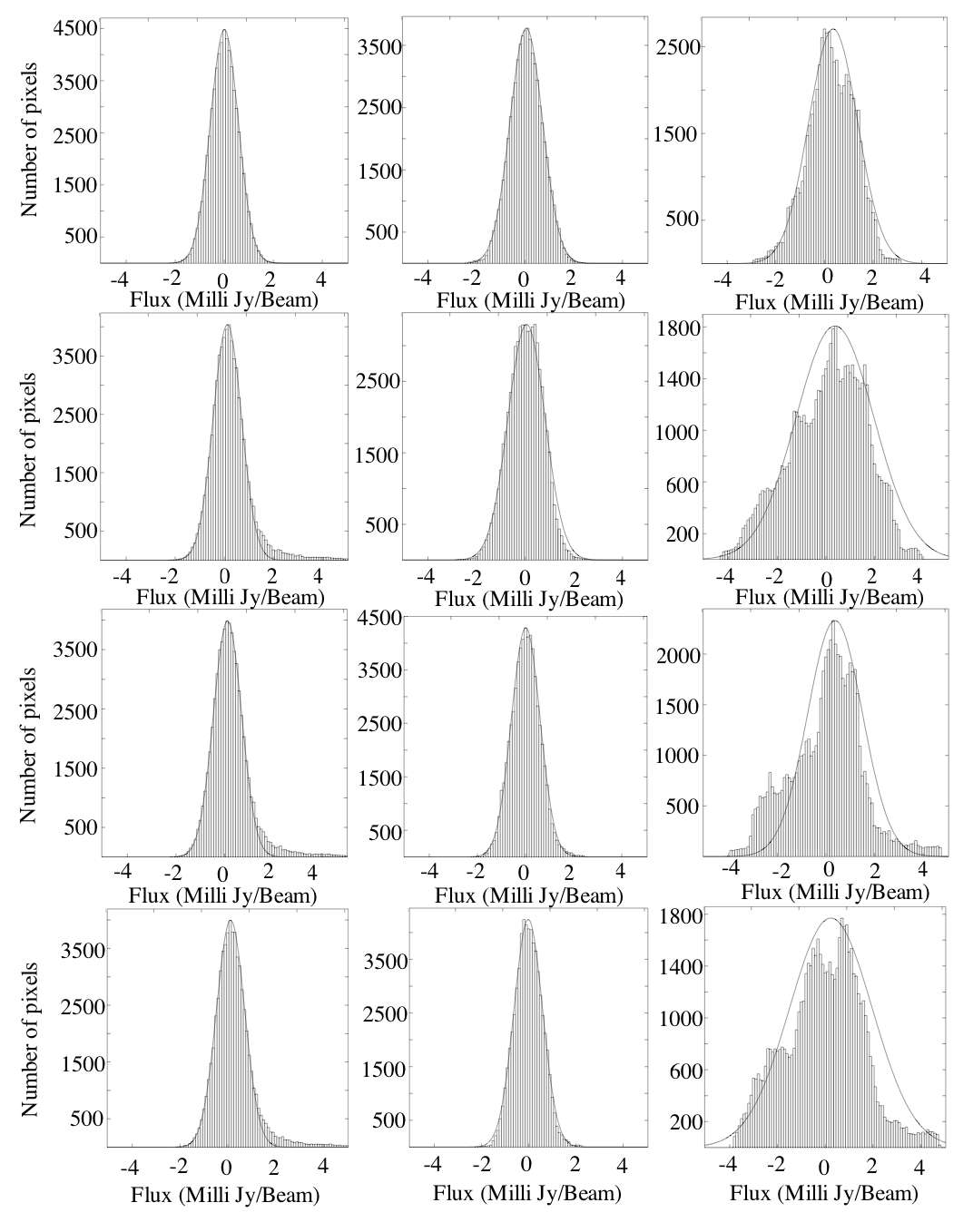}
\caption{Histogram of the intensities in a single channel of
DDO168, B+C+D--configurations. The row at the top is for reference
and represents the distribution in a line--free channel, channel 87,
robust = 0.5, and without any cleaning applied. The first column
corresponds to the full resolution maps, second column to the
15$\arcsec$ residual map and the third column to the 45$\arcsec$
residual map. Rows 2, 3, and  4 are histograms of a line channel,
channel 60, cleaned using MS-CLEAN with different values for
$\alpha$, from top to bottom: $\alpha=0$, $\alpha=0.2$, and
$\alpha=0.4$. Note that the first column corresponds to the final
(cleaned and restored) map which explains the tail to high
intensities; the remaining results are based on residual maps.
\label{figC3}}
\end{figure}

As a final check, we looked into the noise characteristics of MS-CLEAN. In Fig.~\ref{figC3} we present noise histograms for different 
choices of $\alpha$ corresponding to the cleaned maps at full resolution, and the 15$\arcsec$ and 45$\arcsec$ residual maps and 
compare these with noise histograms on a line free channel at the same scales. We see that MS-CLEAN does not affect the nature 
of the noise in any of the maps shown (the tail to positive intensities in rows 2, 3, and 4, reflects genuine emission).

\section{The Data}

\begin{figure}
\epsscale{1.0}
\includegraphics[angle=0,width=1.0\textwidth]{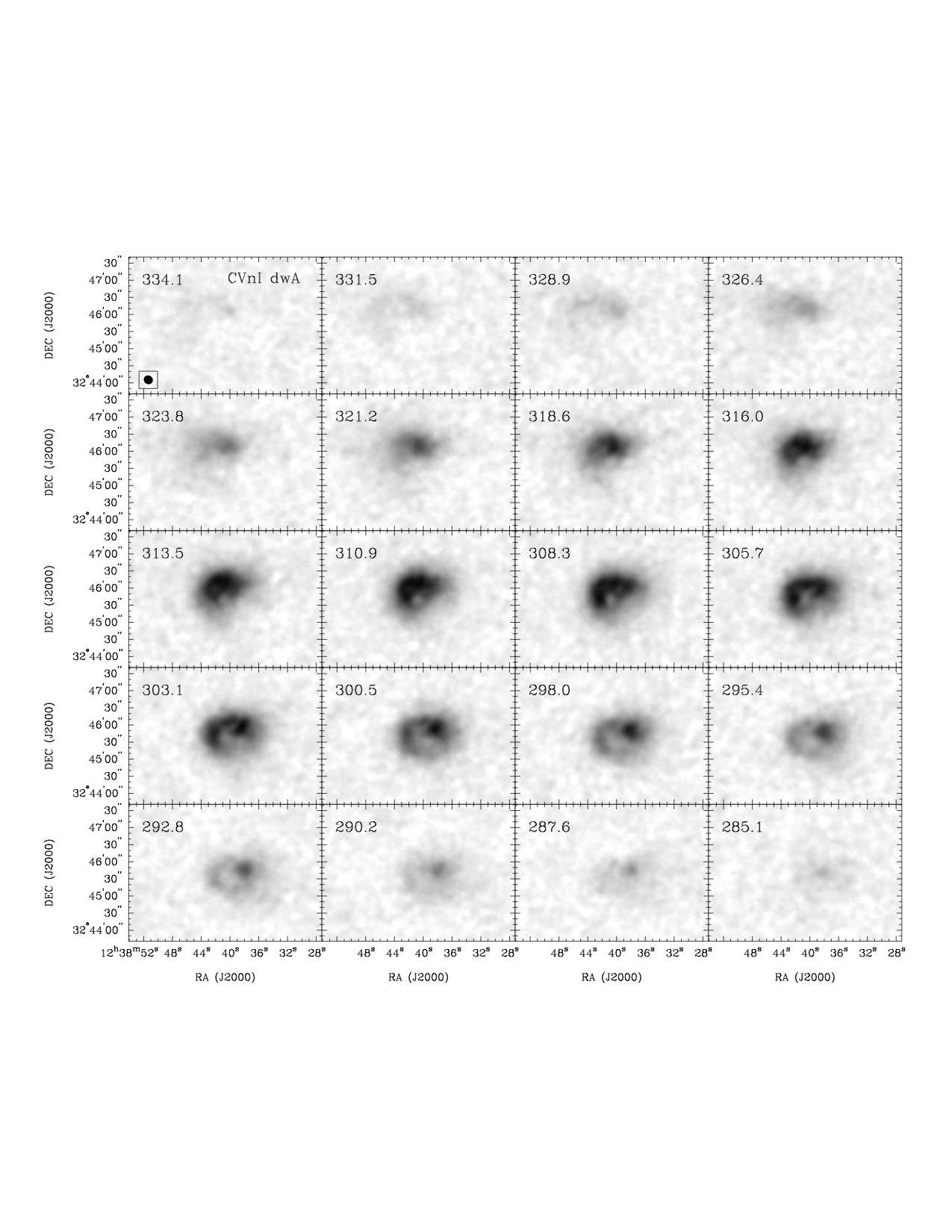}
\caption{Channel maps for CVnIdwA made from the Hanning-smoothed Naturally-weighted \HI-line cube.
The synthesized beam is given in the left bottom corner of the top left panel. In each panel we give
the observed radial velocity in km s$^{-1}$.
\label{fig-cvnidwach}}
\end{figure}

\begin{figure}
\epsscale{1.0}
\includegraphics[angle=0,width=1.0\textwidth]{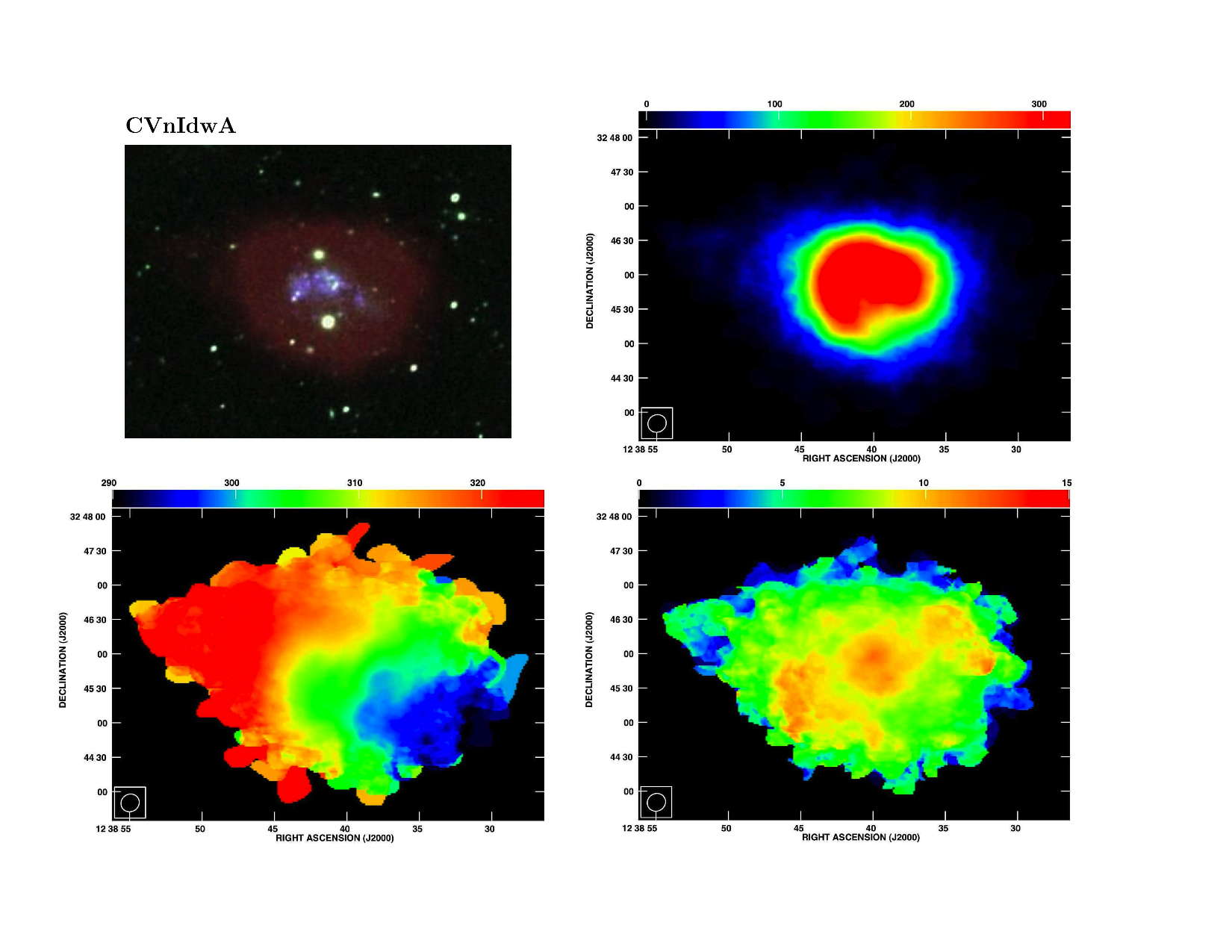}
\caption{Basic data for CVnIdwA. \HI\ data are made from the Hanning-smoothed Naturally-weighted cube. 
{\it Upper left:} False color picture combining \HI\ ({\it red}), $V$-band ({\it green}), and FUV ({\it blue}).
{\it Upper right:} Integrated \HI. The color wedge is in units of Jy beam$^{-1}$ m s$^{-1}$.
{\it Bottom left:} Intensity-weighted velocity field.
{\it Bottom right:} Intensity-weighted velocity dispersion.
The color wedge for both bottom panels is in km s$^{-1}$.
\label{fig-cvnidwacol}}
\end{figure}

We present the basic data for each galaxy in Figures \ref{fig-cvnidwach} to 90.
Figures \ref{fig-cvnidwach} and \ref{fig-cvnidwacol} are shown here, and the rest are available on-line.
Each galaxy has two pages: the channel maps and a combination of images.
The requirement for the channel maps was that each galaxy fit on one page, and so some have
binned pixels and none show every channel. However, we show the full range of channels with emission.
On the image page, the false color image in the upper left combines the integrated \HI\ map (red), $V$ (green), and FUV (blue) 
and allows one to see the relationship of the gas and stars.
The other images are the first three moment maps from the 
Naturally-weighted cube: velocity-integrated \HI, flux-weighted velocity field, and flux-weighted velocity dispersion
for each galaxy.
All of the flux-weighted velocity dispersion maps (moment two) were made with a scale of 0 (dark blue) to 15 km s$^{-1}$ (red).

\clearpage

\clearpage

\clearpage

\clearpage

\clearpage

\includegraphics[angle=0,width=1.0\textwidth]{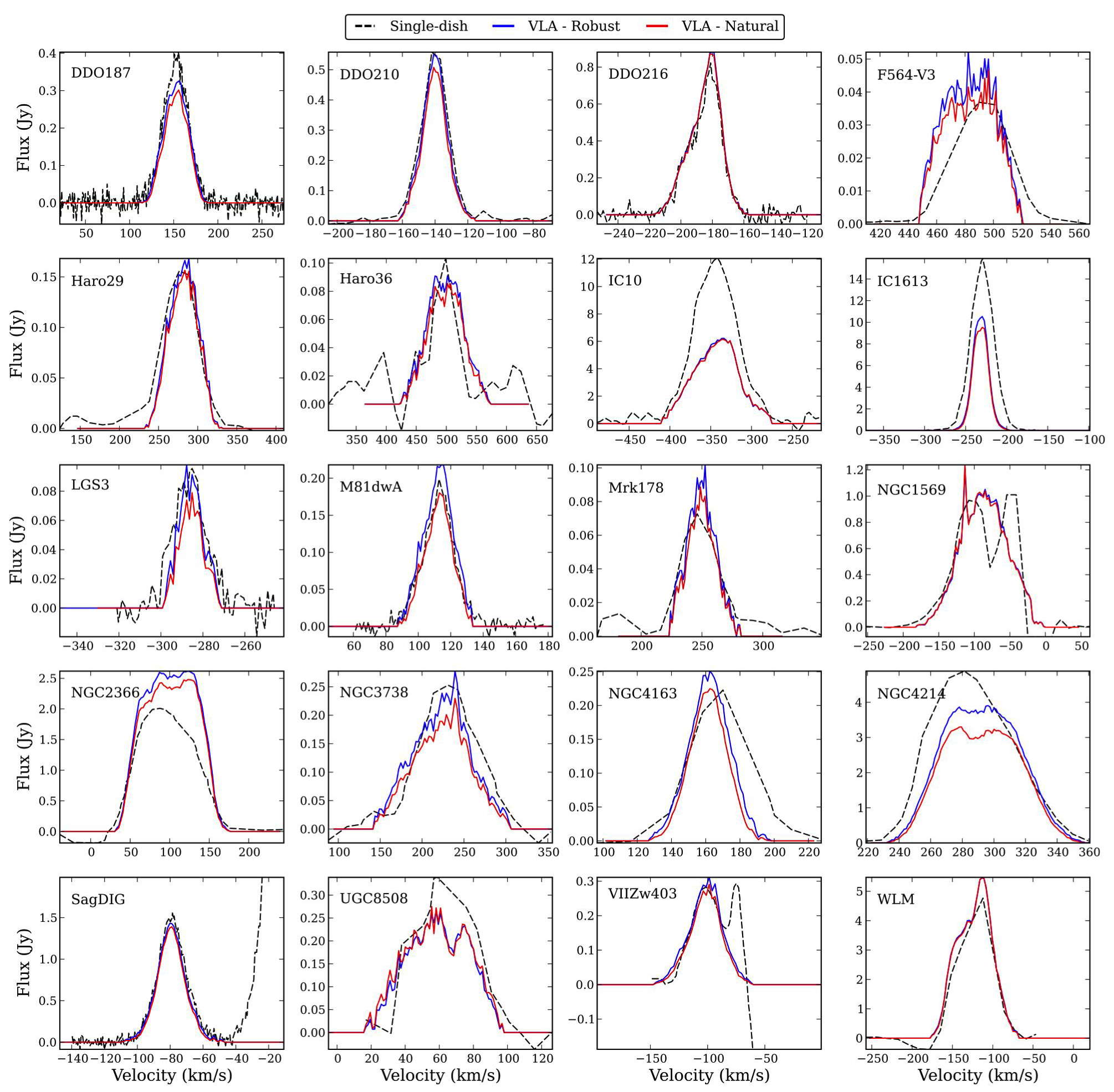}

Figure 4 (continued)

\clearpage

\clearpage

\includegraphics[angle=0,width=0.9\textwidth]{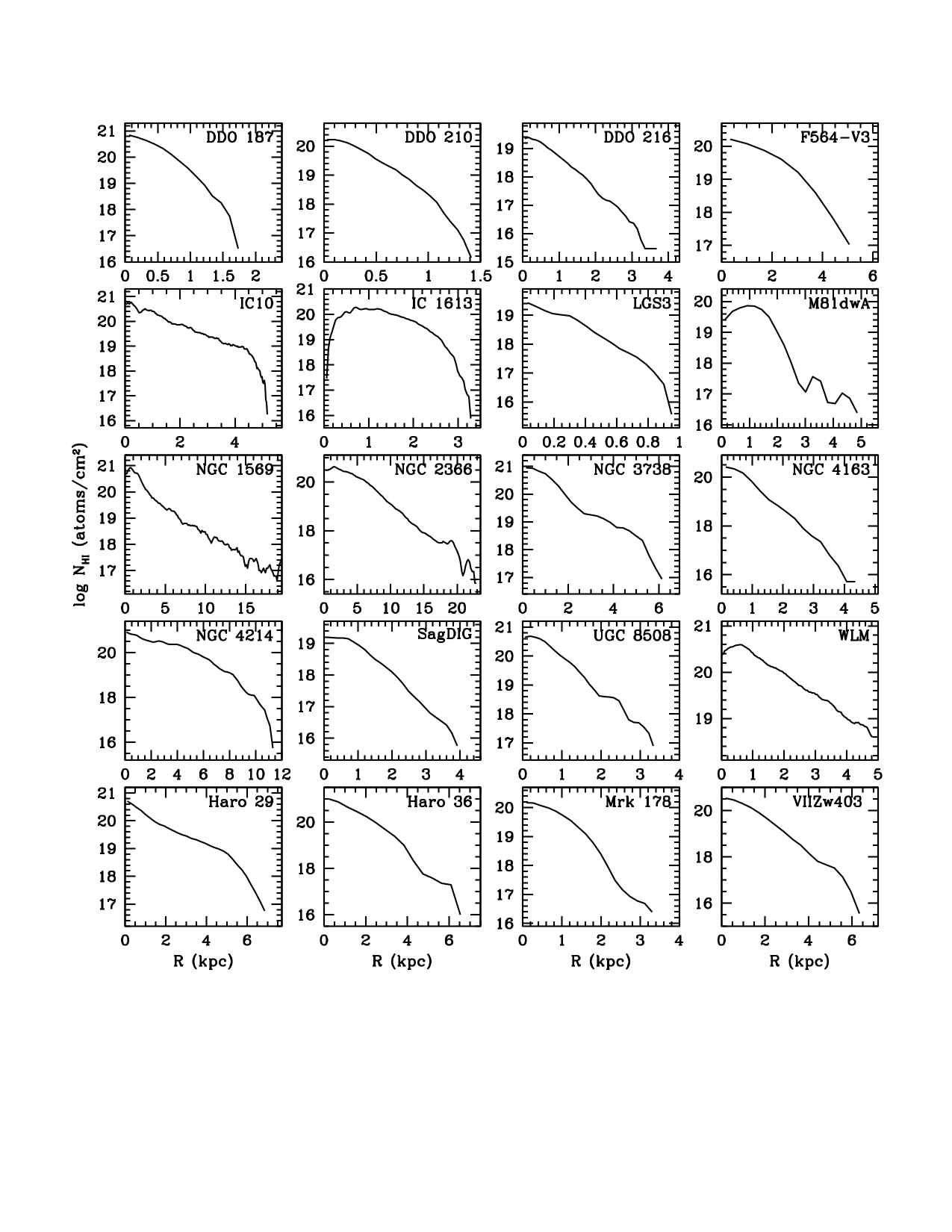}

Figure 5 (continued)

\clearpage

\clearpage

\clearpage

\clearpage

\clearpage

\clearpage

\clearpage

\clearpage

\changetext{0.75in}{1.75in}{}{-1in}{}
\begin{deluxetable}{lccccccccccc}
\tabletypesize{\tiny}
\rotate
\tablenum{1}
\tablecolumns{10}
\tablewidth{689pt}
\tablecaption{The Galaxy Sample \label{tab-sample} }
\tablehead{
\colhead{} & \colhead{} & \colhead{D} & \colhead{} & \colhead{M$_V$}
& \colhead{$R_H$\tablenotemark{c}} & \colhead{$R_D$\tablenotemark{c}} & \colhead{} 
& \colhead{log SFR$_D^{H\alpha}$\tablenotemark{e}} & \colhead{log SFR$_D^{FUV}$\tablenotemark{e}}
& \colhead{} & \colhead{} \\
\colhead{Galaxy}  &  \colhead{Other names\tablenotemark{a}}
& \colhead{(Mpc)} & \colhead{Ref\tablenotemark{b}} & \colhead{(mag)}
& \colhead{(arcmin)} & \colhead{(kpc)} & \colhead{E($B-V$)$_f$\tablenotemark{d}} 
& \colhead{(M\solar\ yr$^{-1}$ kpc$^{-2}$)} & \colhead{(M\solar\ yr$^{-1}$ kpc$^{-2}$)} 
& \colhead{$12+\log {\rm O/H}$\tablenotemark{f}}  & \colhead{Ref\tablenotemark{g}}
}
\startdata
\cutinhead{Im Galaxies}
CVnIdwA   & UGCA 292                                                                          &  3.6 & 1             & -12.4 & 0.87 & $0.57 \pm 0.12$ & 0.01 & $-2.58 \pm 0.01$ & $-2.48 \pm 0.01$ & $7.3 \pm 0.06$ & 24 \\
DDO 43     &  PGC 21073, UGC 3860                                                  &  7.8 & 2              & -15.1 & 0.89 & $0.41\pm 0.03$ & 0.05 & $-1.78 \pm 0.01$ & $-1.55 \pm 0.01$ & $8.3 \pm 0.09$ & 25 \\
DDO 46     &  PGC 21585, UGC 3966                                                  &  6.1 & \nodata  & -14.7 & \nodata & $1.14 \pm 0.06$ & 0.05 & $-2.89 \pm 0.01$ & $-2.46 \pm 0.01$ & $8.1 \pm 0.1$  & 25 \\
DDO 47     &  PGC 21600, UGC 3974                                                  &  5.2  & 3             & -15.5 & 2.24 & $1.37 \pm 0.06$ & 0.02 & $-2.70 \pm 0.01$ & $-2.40 \pm 0.01$ & $7.8 \pm 0.2$ & 26 \\
DDO 50     & PGC 23324, UGC 4305, Holmberg II, VIIZw 223        &  3.4 & 1             & -16.6 & 3.97 & $1.10 \pm 0.05$ & 0.02 & $-1.67 \pm 0.01$ & $-1.55 \pm 0.01$ & $7.7 \pm 0.14$ & 27 \\
DDO 52     &  PGC 23769, UGC 4426                                                  & 10.3 & 4            & -15.4 & 1.08 & $1.30 \pm 0.13$ & 0.03 & $-3.20 \pm 0.01$ & $-2.43 \pm 0.01$ & (7.7)                   & 28 \\
DDO 53     &  PGC 24050, UGC 4459, VIIZw 238                              &  3.6 & 1             & -13.8 & 1.37 & $0.72 \pm 0.06$ & 0.03 & $-2.42 \pm 0.01$ &$-2.41 \pm 0.01$ & $7.6 \pm 0.11$ & 27 \\
DDO 63     &  PGC 27605, Holmberg I, UGC 5139,  Mailyan 044   &  3.9 & 1             & -14.8 & 2.17 & $0.68 \pm 0.01$ & 0.01 & $-2.32 \pm 0.01$ & $-1.95 \pm 0.00$ & $7.6 \pm 0.11$ & 27 \\
\enddata
\end{deluxetable}

\clearpage

\changetext{-0.75in}{-1.75in}{}{1in}{}
\begin{deluxetable}{lcll}
\tabletypesize{\scriptsize}
\rotate
\tablenum{2}
\tablecolumns{4}
\tablewidth{0pt}
\tablecaption{The Data Sets \label{tab-datasets}}
\tablehead{
\colhead{Data set} & \colhead{No. Galaxies} & \colhead{Purpose} & \colhead{Status} \\
}
\startdata
VLA HI-line maps\tablenotemark{a}           &     & Trace cloud formation                                  &  \\
\hspace{0.15truein} Archives (1.3 km s$^{-1}$) &  9 &                                                                          & This paper \\
\hspace{0.63truein} (2.6 km s$^{-1}$)                  & 11 &                                                                         & This paper \\
\hspace{0.15truein} New Obs (1.3 km s$^{-1}$) & 11 &                                                                         & This paper \\
\hspace{0.65truein} (2.6 km s$^{-1}$)                   & 10 &                                                                         & This paper \\
H$\alpha$ images                                       & 41 & Trace current star formation                       & Hunter \& Elmegreen (2004) \\
{\it GALEX} UV                                              &       & Trace recent star formation                       & \\
\hspace{0.15truein} Archives                     & 27 &                                                                         & Hunter et al.\ (2010), Herrmann et al.\ (2012) \\
\enddata
\tablenotetext{a}{1.3 km s$^{-1}$ and 2.6 km s$^{-1}$ refer to the channel separation.}
\end{deluxetable}

\clearpage

\changetext{0.75in}{}{}{}{}
\begin{deluxetable}{lccccccccccccc}
\tabletypesize{\tiny}
\rotate
\tablenum{3}
\tablecolumns{14}
\tablewidth{0pt}
\tablecaption{The VLA Observations \label{tab-obs}}
\tablehead{
\colhead{} & \colhead{} & \colhead{} & \colhead{}
& \colhead{} & \colhead{} & \colhead{}
& \multicolumn{3}{c}{{\sc robust}-weighted map cube} 
& \colhead{}
& \multicolumn{3}{c}{Natural-weighted map cube} \\
\cline{8-10} \cline{12-14}
\colhead{}  & \colhead{} & \colhead{}
& \colhead{} & \colhead{TOS\tablenotemark{c}} & \colhead{Ch sep}  & \colhead{Final} 
& \colhead{Beam\tablenotemark{d}} & \colhead{Beam PA} & \colhead{RMS}
& \colhead{}
& \colhead{Beam\tablenotemark{d}} & \colhead{Beam PA} & \colhead{RMS} \\
\colhead{Galaxy}  & \colhead{Array} & \colhead{Obs Dates\tablenotemark{a}}
& \colhead{Project ID\tablenotemark{b}} & \colhead{(hr)} & \colhead{(km s$^{-1}$)} & \colhead{\# Ch}  
& \colhead{(arcs$\times$arcs)} & \colhead{(deg)} & \colhead{(mJy bm$^{-1}$)}
& \colhead{}
& \colhead{(arcs$\times$arcs)} & \colhead{(deg)} & \colhead{(mJy bm$^{-1}$)}
}
\startdata
\cutinhead{Im Galaxies}
CVnIdwA   &  B     & 07 Dec 3, 08 Feb 6                          & AO215,AH927 &  8.3 & 1.3  & 101 & 10.9$\times$10.5 &  82.0 &  0.65  & &16.3$\times$15.1 & -57.7 &  0.58 \\
                    & CnB & 08 Feb 7                                            & AH927               &  0.9  &         &         &                                    &           &            & &                                  &           &           \\
                    & C      &  99 Jan 24, 08 Jul 11                      & AL471,AO215  &   6.5 &        &         &                                    &           &            & &                                  &           &           \\
                    & D      & 08 Jul 11, 08 Jul 31, 08 Aug 16    & AO215,AH927 &   2.6 &        &         &                                    &           &            & &                                  &           &           \\
DDO 43     & B       & 08 Jan 13, 08 Jan 22, 08 Jan 26  & AH927               & 10.9 & 2.6 & 106  & 8.1$\times$6.0      &   88.5 & 0.55  & &  14.3$\times$12.3  &-79.7 & 0.50  \\
                    & C       &  93 Jul 5, 00 Apr 9                           & AS506,AH695 &   4.6 &        &         &                                    &           &            & &                                  &           &           \\
                    & D       &  96 Sep 20, 01 Oct 19                     & AS591,AH753 &   2.9 &        &         &                                    &           &            & &                                  &           &           \\
DDO 46     &  B       & 07 Dec 19,08 Jan 6,08 Jan 15,08 Jan 16 & AH927 & 12.2 & 2.6 & 102 &   6.3$\times$5.2      & 81.5  & 0.57   & &   9.8$\times$7.1    &  73.9 & 0.52 \\
\enddata
\tablenotetext{a}{Date when the observations began.}
\tablenotetext{b}{AH927 is the LITTLE THINGS project.}
\tablenotetext{c}{Time on Source (TOS).}
\tablenotetext{d}{Beam major $\times$ minor axis.}
\end{deluxetable}

\clearpage

\changetext{-0.75in}{}{}{}{}
\begin{deluxetable}{lccc}
\tabletypesize{\small}
\tablenum{4}
\tablecolumns{4}
\tablewidth{0pt}
\tablecaption{Integrated Masses \label{tab-masses}}
\tablehead{
\colhead{}  
& \colhead{Conversion Factor\tablenotemark{a}} & \colhead{log $S_\nu$\tablenotemark{b}} 
& \colhead{log $M_{HI}$} \\
\colhead{Galaxy}  
& \colhead{(M\solar (Jy beam$^{-1}$ m s$^{-1}$)$^{-1}$)} & \colhead{(Jy beam$^{-1}$ m s$^{-1}$)} 
& \colhead{(M\solar)} 
}
\startdata
\cutinhead{Im Galaxies}
CVnIdwA  &   52.98 &  5.95 &  7.67 \\
DDO 43   &  595.15 &  5.45 &  8.23 \\
DDO 46   &  530.30 &  5.55 &  8.27 \\
DDO 47   &  134.65 &  6.46 &  8.59 \\
DDO 50   &   125.49 &  6.75 &  8.85 \\
DDO 52   &  1404.03 &  5.28 &  8.43 \\
DDO 53   &   169.40 &  5.49 &  7.72 \\
DDO 63   &   152.10 &  6.01 &  8.19 \\
\enddata
\tablenotetext{a}{Factor to convert flux in the {\sc robust}-weighted integrated \HI\ map to mass in solar masses.
Note this conversion factor contains the necessary division by the number of pixels per beam, as
discussed in \S 4.}
\tablenotetext{b}{Integrated flux measured on the integrated \HI\ map made from the {\sc robust}-weighted image cube.}
\end{deluxetable}

\clearpage

\changetext{0.5in}{0.5in}{}{-0.5in}{}
\begin{deluxetable}{lllp{18cm}}
\tabletypesize{\tiny}
\rotate
\tablenum{5}
\tablecolumns{4}
\tablewidth{0pt}
\tablecaption{Velocity Profiles and Single-dish comparison \label{tab-velprofilecomp}}
\tablehead{
\colhead{Galaxy}  & \colhead{Instrument} & \colhead{Ref} & \colhead{Comments} 
}
\startdata
\cutinhead{Im Galaxies}
CVnIdwA & Green Bank 42m & 13 & Excellent agreement. \\ 
DDO 43 & Green Bank 91m & 13 & Profiles match very well, with the VLA only missing a small part of single-dish emission near the center. \\
DDO 46 & Green Bank 91m & 8,13 & Profiles match quite well, with only minimal difference in the center. \\
DDO 47 & Green Bank 42m & 13 & Excellent agreement. \\
DDO 50 & Green Bank 91m & 9 & The shapes are roughly the same, but the single dish picks up more emission on the high velocity side. This could result from gas with low surface brightness. \\
DDO 52 & Green Bank 42m & 13 & Excellent agreement.  \\
DDO 53 & Effelsberg 100m & 5 & Only a partial single-dish spectrum available, due to Galactic confusion. The VLA profile matches the available data quite well.  \\
DDO 63 & Effelsberg 100m & 13 & Good agreement, with the VLA recovering ~10\% less emission at the peak. This falls within the measurement uncertainty. \\
\enddata
\tablerefs{(1) Giovanelli et al. 2007, (2) Eder \& Schombert 2000, (3) Huchtmeier \& Seiradakis 1985, (4) Huchtmeier \& Richter 1986, (5) Huchtmeier et al. 2003, (6) Hunter \& Gallagher 1985, (7) Koribalski et al. 2004, (8) Lewis, Helou, Salpeter 1985, (9) Rots 1980, (10) Schneider et al. 1990, (11) Schneider et al. 1992, (12) Shostak \& Allen 1980, (13) Springob et al. 2005, (14) Thuan \& Martin 1981, (15) Young \& Lo 1997.  DDO 69 and DDO 70 show data that was previously unpublished, but taken by Young and processed in the same manner as the published SagDIG data.}
\end{deluxetable}

\clearpage

\changetext{-0.5in}{-0.5in}{}{0.5in}{}
\begin{deluxetable}{lccccc}
\tabletypesize{\small}
\tablenum{6}
\tablecolumns{6}
\tablewidth{0pt}
\tablecaption{Surface Density Photometry Parameters\tablenotemark{a} \label{tab-sd}}
\tablehead{
\colhead{}  
& \colhead{R.A. (2000.0)} & \colhead{Decl. (2000.0)} & \colhead{P.A.\tablenotemark{b}} & \colhead{} & \colhead{Incl}  \\

\colhead{Galaxy}  
& \colhead{(hh mm ss.s)} & \colhead{(dd mm ss)} & \colhead{(deg)} & \colhead{$b/a$\tablenotemark{c}} & \colhead{(deg)} 
}
\startdata
\cutinhead{Im Galaxies}
CVnIdwA  &  12 38 40.2  &  +32 45 40 & 80    & 0.78 & 41.0  \\
DDO 43     &   7 28 17.8   &  +40 46 13 &   6.5 & 0.70 & 48.5  \\
DDO 46     &   7 41 26.6   &  +40 06 39 & 84    & 0.89 & 28.6  \\
DDO 47     &   7 41 55.3   & +16 48 08  & -70   & 0.51 & 64.4  \\
DDO 50     &   8 19 08.7   & +70 43 25  & 18    & 0.72 & 46.7  \\
DDO 52     &   8 28 28.5   &  +41 51 21 &   5    & 0.67 & 51.1  \\
DDO 53     &   8 34 08.0   &  +66 10 37 &  81   & 0.51 & 64.4  \\ 
DDO 63     &   9 40 30.4   &  +71 11 02 &  0     & 1.00 &   0.0   \\
\enddata
\tablenotetext{a}{From the $V$-band (Hunter \& Elmegreen 2006).}
\tablenotetext{b}{Position Angle (P.A.).}
\tablenotetext{c}{Minor-to-major axis ratio as measured in an outer $V$-band surface brightness isophote.
For computing the inclination of the galaxy, the intrinsic $b/a$ is assumed to be 0.3 for dwarf galaxies
(Hodge \& Hitchcock 1966, van den Bergh 1988).}
\end{deluxetable}

\end{document}